\documentclass[useAMS,usenatbib]{mn2e}
\usepackage{graphicx,epsfig,lscape}
\usepackage{times,deluxetable}
\title[An unbiased sample of bright CSS and GPS Sources]{An unbiased sample of bright southern Compact Steep Spectrum and Gigahertz Peaked Spectrum Sources}
\author[K. E. Randall, A. M. Hopkins, R. P. Norris \& P. G. Edwards]{K. E. Randall$^{1,2}$\thanks{E-mail:
krandall@physics.usyd.edu.au}, A. M. Hopkins$^{3}$, R. P. Norris$^{2}$ and P. G. Edwards$^{2}$\\
$^{1}$Sydney Institute for Astronomy, A28, School of Physics, The University of Sydney, Camperdown, NSW 2006, Australia\\
$^{2}$CSIRO Astronomy and Space Science, P.O. Box 76, Epping, NSW
1710, Australia\\
$^{3}$Australian Astronomical Observatory, PO Box 296, Epping, NSW 1710, Australia\\}
\begin{document}

\date{Accepted  . Received ; in original form 2010 October 1}

\pagerange{\pageref{firstpage}--\pageref{lastpage}} \pubyear{2011}

\maketitle

\label{firstpage}
\begin{abstract}
Compact Steep Spectrum (CSS) and Gigahertz Peaked Spectrum (GPS) sources are classes of compact, powerful, extragalactic objects. These sources are thought to be the earliest stages in the evolution of radio galaxies, capturing the ignition (or, in some cases, re-ignition) of the AGN. As well as serving as probes of the early stages of large-scale radio sources, these sources are good, stable, amplitude calibrators for 
radio telescopes.  We present an unbiased flux density limited ($>1.5$\,Jy at 2.7\,GHz) catalogue of these objects in the Southern Hemisphere, including tabulated data, radio spectra, and 
where available, optical images and measurements. The catalogue contains 26 sources, consisting of 2 new candidate and 15 known CSS sources, and 9 known GPS sources. We present new Australia Telescope Compact Array (ATCA) data on ten of these 26 sources, and data on a further 42 sources which were excluded from our final sample. This bright sample will serve as a reference sample for comparison with subsequent faint (mJy level) samples of CSS and GPS candidates currently being compiled.
\end{abstract}


\section{Introduction}
\label{int}
Gigahertz Peaked Spectrum (GPS) and Compact Steep Spectrum (CSS) Sources are strong, compact radio sources named for the shape of their radio spectra. CSS sources have a steep ($\alpha\leq-0.5$)\footnote{We define $\alpha$ as $S\propto\nu_{o}^{\alpha}$, where S is flux density and $\nu_{o}$ is the observer's frame frequency. We use $\nu_{o}$ for the frequency in the observer's frame and $\nu_{r}$ to denote rest frame frequency throughout this paper.} power-law spectrum which continues to 1 GHz or below, while GPS sources reach a peak flux density at a frequency around 1 GHz. Approximately 10\% and 30\% of bright centimetre wavelength radio sources are GPS or CSS objects respectively \citep{Review}. Catalogued CSS and GPS sources have similar radio powers, L$\approx$10$^{25}$\,WHz$^{-1}$ \citep{NatCSSs,Review,cfanti}, have low fractional polarization \citep{Review}, and neither class exhibits strong variability ($\geq$10\% over a year) in general. However,  GPS sources, particularly GPS quasars, tend to be more variable than CSS sources \citep{snell}.

CSS and GPS sources are generally unresolved in low resolution imaging, as GPS sources are typically $<$1\,kpc, and CSS sources are one to several tens of kpc in size \citep{Review,faintgps}. High resolution radio imaging reveals structures similar to Fanaroff-Riley Type I and II galaxies \citep{fr}, such as lobes or jets on either side of a compact core (the core itself may not be detected), and sources with asymmetrical emission, which are attributed to relativistic beaming \citep{tasso}.

The turnover in GPS spectra (and, at lower frequencies, CSS spectra) is thought to be caused by synchrotron self absorption, although free-free absorption may play a role \citep{cfanti}.

Because of their small sizes, and spectra, CSS and GPS sources are widely thought to represent the start of the evolutionary path to large-scale radio sources \citep{young,tinti,cfanti}. In this model, GPS sources expand and evolve into  CSS sources \citep{Review,faintgps,marecki}, which in turn expand and evolve into the largest radio sources, FRI/II galaxies \citep{fr,evolution}. An alternative model is that rather than being young, they may be ``frustrated'', their small size being caused by a dense medium which confines the lobes \citep{oldodea}, and their steep spectrum being caused by ageing electrons. Further, \citet{rfanti} has suggested that some CSS or GPS sources are prematurely dying radio sources in which the energy supply has ceased, leaving only diffuse emission and a steep spectrum core.\\

These sources are an ideal resource for investigation of galaxy evolution and formation, as well as AGN feedback, given that the ``youth" scenario seems highly plausible. Not only may they be young AGN but they also have star formation occurring due to interactions and mergers \citep{Review,labiano,morganti}.
In addition bright CSS sources are invaluable as flux calibrators for
current and next-generation radio-telescopes, such as the Australian Square Kilometre Array Pathfinder \citep[ASKAP;][]{askap}.

Although CSS and GPS sources are relatively common in radio surveys of strong sources, there are few complete samples, making it difficult to characterise the statistical properties of the population as a whole. In this paper we compile an unbiased flux density limited sample ($>1.5$\,Jy at $\nu_{o}=2.7$\,GHz) of bright CSS/GPS sources and in a subsequent paper we will compile a sample extending to much fainter flux densities.

We discuss the selection of the bright CSS and GPS candidates in Section~\ref{sec:selection}, and details of new observations made using the ATCA are listed in Section~\ref{sec:obs}. The resultant catalogue is presented and explored in Section~\ref{sec:results} and the analysis of our results is discussed in Section~\ref{sec:discussion}. Our conclusions are presented in Section~\ref{sec:concl}. Throughout this analysis, we use the cosmological parameters, $\Omega_{M}=0.27$, $\Omega_{\Lambda}=0.73$ and $H_0=71$\,kms$^{-1}$Mpc$^{-1}$. All frequencies and luminosities are in the observer's frame, indicated by $\nu_{o}$ unless otherwise stated.

\section{Candidate Selection}
\label{sec:selection}

Our selection process is aimed at constructing a uniform sample of CSS sources. We note that the selection of GPS candidates is inherently difficult, and requires data over a broad wavelength range. Contamination from variable objects (such as quasars) can produce spectra which may be falsely identified as GPS sources \citep[e.g.,][]{hancock}. We did not specifically search for GPS sources and the sources catalogued here as GPS sources were found as a by-product of the CSS candidate search. All are previously known GPS sources. This study will be missing sources with spectral peaks at higher frequencies than our selection frequency or which are otherwise excluded by our CSS selection criteria.
\subsection{Initial selection criteria}
\label{sec:crit}
In Table~\ref{table:selection} we list the criteria used to select candidates, and also summarize the number of sources remaining after each selection criterion was applied. The initial selection criteria are above the horizontal line, whilst the secondary selection criteria are below this line. Our selection begins with the Parkes 1990 Catalogue \citep[PKSCAT90;][]{PKScat} which is a complete catalogue at 2.7 GHz\citep{Bolt}, covering the frequency range 80\,MHz to 22\,GHz, containing 8264 radio sources. 

\begin{figure}
\begin{center}
\includegraphics[scale=0.4]{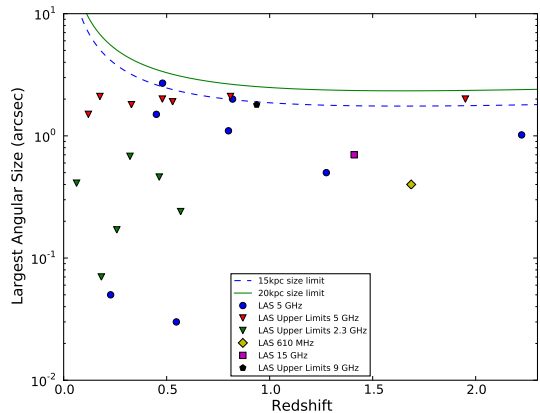} 
\caption{Largest Angular Size (LAS) versus redshift for our sample of 26, with LAS references given in Table~\ref{table:las}.}
\label{fig:angsize}
\end{center}
\end{figure}

\begin{table*}
\begin{center}
\caption{\textsc{Summary of sources remaining after each initial and secondary selection criterion was applied}}
\label{table:selection}
\begin{tabular}{|ccc|} \hline
Criterion Number & Selection Criterion & Sources Remaining\\
\hline
0 & PKSCAT90 & 8264\\
1 & Four flux density measurements in PKSCAT90, each at a different frequency. & 2958\\
 & This gives three two-point spectral index measurements using adjacent frequencies & \\
2 & A flux density $S_{2.7}>1.5$\,Jy at $\nu_{o}=2.7$\,GHz & 373\\
3 & $\delta<0^{\circ}$,as most possible CSS or GPS sources in the North are already catalogued & 252\\
4 & $|\textit{b}|>10^{\circ}$, to avoid confusing galactic plane sources & 218\\
5 &$\alpha^{5}_{2.7}<-0.5$, selection spectral index used to select CSS and GPS sources  & 115\\
6 & LAS$<30''$ in NVSS/SUMSS & 77\\
7 & Has a flux density measured in the PMN Surveys & 76\\
\hline
8 & Steep across spectrum ($\alpha_{fit}<-0.5$) & 62\\
& ($>0.03$ scatter in rms of fit) and not flat-spectrum, a H{\sc ii} region, or SNR& \\
9 & AT20G catalogue extended tag (LAS$<10''$) & 41\\
10a & ATCAL (flux emission on scales $<8''$) & 35\\
10b & LAS in literature ($<8''$) & 35\\
11 & Physical Size Cut of 20\,kpc & 26\\
\hline
\end{tabular}
\end{center}
\end{table*}
\subsubsection{Selection criteria 1-4: Bright radio sources}
From PKSCAT90, we used the first four selection criteria (Table~\ref{table:selection}) to select bright southern radio sources (outside the Galactic plane), with multiple flux density measurements. Of the 8264 sources from PKSCAT90, 218 remained after these four steps. 
\subsubsection{Selection criterion 5: Steep radio spectra}
We applied a steep spectral index cut to the 218 sources, using the spectral index between 2.7 and 5\,GHz, of $\alpha^{5}_{2.7}<-0.5$. This is analogous to those used to select candidates from other surveys \citep[and references therein]{Review}, and selected 115 sources. Two previous CSS candidates, J1424$-$4913 and J2206$-$1835 \citep{PE}, do not appear as initial candidates as both have a spectral index $\alpha^{5}_{2.7}>-0.5$.
\subsubsection{Selection criterion 6: Compact sources}
To ensure that the remaining 115 steep spectrum sources are compact, we placed a Largest Angular Size (LAS) limit of $<30''$ on our candidates using angular size measurements from the Sydney University Molonglo Sky Survey \citep[SUMSS;][]{SUMSS1,SUMSS2} or the NRAO VLA Sky Survey \citep[NVSS;][]{NVSS}. NVSS covers the sky down to a declination of $\delta=-40^{\circ}$ and SUMSS extends from a declination of $\delta=-90^{\circ}$ to $\delta=-30^{\circ}$. 

If the source had neither a catalogued SUMSS or NVSS source within $30''$ of the PKSCAT90 radio position, we first checked the accuracy of the radio position (as the PKSCAT90 radio positions have large uncertainties, $<20''$) by using the SUMSS and or NVSS image postage stamp servers to confirm the location of the source. These images were also used to check whether the source was extended. If it was extended, or resolved (i.e. a radio double or triple source, or a complex source), the source was removed from the list of candidates. Neither SUMSS nor NVSS give flux density measurements for very extended or complex radio sources. We used NED to confirm the radio position of these non-SUMSS or NVSS sources, and we were able to remove 24 non-matches by checking the available radio images and references confirming that these sources were extended\footnote{\citet{ext5,Cul3,ext7,ext8,ext2,ext3,morgsize,ext10,ext9,kapahi,reid,ext11,ext12}}.
\begin{table}
\caption{\textsc{Summary of Surveys or Catalogue data for the final sample}}
\label{table:cats}
\begin{tabular}{|cccc|} \hline
Survey/Catalogue & Frequency & Number & Reference \\
& $\nu_{o}$, GHz & of Sources & \\
\hline
\footnotesize PKSCAT90 1990 & \footnotesize 0.08, 0.178, 0.635,  & \footnotesize 26 & 1\\
 & 1.4, 2.7, 5, 8.4, 22 & &\\
\footnotesize PMN & \footnotesize 4.85 & \footnotesize 26 & 2\\
\footnotesize MRC & \footnotesize 0.408 & \footnotesize 26 & 3\\
\footnotesize Culgoora & \footnotesize 0.08, 0.160 & \footnotesize 14 & 4\\
\footnotesize SUMSS & \footnotesize 0.843 & \footnotesize 15 & 5, 6\\
\footnotesize NVSS & \footnotesize 1.4 & \footnotesize 15 & 7\\
\footnotesize VLSS & \footnotesize 0.074 & \footnotesize 12 & 8\\
\footnotesize Texas & \footnotesize 0.365 & \footnotesize 13 & 9\\
\footnotesize AT20G & \footnotesize 20 & \footnotesize 26 & 10, 11 \\
\hline
\end{tabular}
\textsc{References.}-- (1) Wright \& Otrupcek 1990, (2) Griffith \& Wright 1993, (3) Large et al. 1981, (4) Slee \& Higgins 1973, 1975, 1977, (5) Bock et al. 1999, (6) Mauch et al. 2003, (7) Condon et al. 1998, (8) Cohen et al. 2007, (9) Douglas et al. 1996, (10) Massardi et al. 2008, (11) Murphy et al. 2010.
\end{table}

\subsubsection{Selection criterion 7: PMN Positions and flux density measurements}
As a final step, each source was checked to see if there was a PMN \citep{PMN} flux density and position, as the PMN survey has more robust flux densities and radio positions than the original PKSCAT90 measurements. One source was removed as it lacked a PMN counterpart, leaving an initial 76 candidate CSS sources. 

\subsection{Secondary Selection}
\label{sec:secondary}
To construct a reliable radio spectrum the flux density information from PKSCAT90 and PMN was augmented with data from the catalogues or surveys listed in Table~\ref{table:cats}. The flux density data for each source ranges from 408\,MHz up to 20\,GHz. For 18 of the 26 sources this range extends down to 74 or 80\,MHz, and this is reflected in our catalogue. Of the 76 CSS candidate sources, 55 have radio data from recent radio observations, described in Section~\ref{sec:obs}. These data were also included in the spectra for the individual sources.

\subsubsection{Selection Criterion 8: Steep Spectra}
We fitted a least-squares power-law to the overall radio spectrum for each source, at the frequencies we had measured flux densities for. We define $\alpha_{fit}$ to be the resultant overall spectral index from this fit. For a source to remain as a candidate, it needed to satisfy a further two requirements. First, it must have $\alpha_{fit}<-0.5$ and second, the sum of the residuals of the power-law fit (in logarithmic units) divided by the number of data points must be less than 0.03, to ensure little scatter in the overall radio spectrum, and to minimize the chances of selecting variable sources. For GPS candidates, those sources which displayed a classical spectral turnover in their radio spectrum, the $\alpha_{fit}<-0.5$ criterion needed to be satisfied above the spectral peak, and a quadratic fit was performed to ensure the sum of the residuals divided by the number of points was less than 0.03 for the available radio spectrum. Fourteen sources were removed at this point, leaving 62 candidates, consisting of 50 CSS and 12 GPS sources. We note that 11 of these GPS sources found as a by-product of the CSS source selection are known and confirmed GPS sources. The previously unknown GPS candidate was later removed (\S~\ref{sec:physical}).
\subsubsection{Selection Criterion 9: AT20G catalogue extended tag}
\label{sec:at20g}
The AT20G catalogue \citep{at20gcat} lists an extended tag for each source, which was used to eliminate any that were not compact. The AT20G survey has a resolution of $10''$, and only sources which were unresolved at this level were kept, resulting in a list of 41 candidates. 
\begin{table*}
\begin{center}
\caption{\textsc{Comparison of selection criteria and sample for O'Dea's sample and our sample}}
\label{table:odea}
\begin{tabular}{|ccc|} \hline
 & O'Dea Sample & This Paper\\
\hline
Selection Frequency &$\nu_{o}=178$\,MHz \& $\nu_{o}=2.7$\,GHz (CSS sources), $\nu_{o}=5$\,GHz (GPS Sources) & $\nu_{o}=2.7$\,GHz\\
Flux Limit & $S_{\nu_{o}=178\,MHz}>10$\,Jy (CSS sources), $S_{\nu_{o}=5\,GHz}>1$\,Jy (GPS sources) & $S_{\nu_{o}=2.7\,GHz}>1.5$\,Jy\\
Power Cut & $>10^{26.75}$ WHz$^{-1}$ at $\nu_{o}=178$\,MHz (CSS sources) & N/A\\
Spectral Index Cut & $\alpha^{5}_{2.7}<-0.5$ for CSS sources selected at $\nu_{o}=2.7$\,GHz & $\alpha^{5}_{2.7}<-0.5$\\
& $\alpha<-0.5$ above the spectral peak for GPS sources & \\
Physical or Angular Size Cut & $>20$\,kpc (CSS sources) & $8''$, then 20\,kpc\\
Peak Frequency Cut & Between 0.4 and 6\,GHz (GPS sources) & N/A\\
Number of Sources & 67 (34 CSS, 33 GPS) & 26 (17 CSS, 9 GPS)\\
Number of objects in common & 8 & 8\\
Number not selected by our criteria & 4 & N/A\\
Number of QSOs & 19 CSS, 14 GPS & 6 CSS, 2 GPS\\
Number of Galaxies & 14 CSS, 19 GPS & 11 CSS, 7 GPS\\
\hline
\end{tabular}
\textsc{Notes.}-- All frequencies are in the observer's frame.
\end{center}
\end{table*}

\subsubsection{Selection Criteria 10a,b: Angular Size}
\label{sec:atcal}
The data from these 41 objects were further investigated using the Australia Telescope Calibrator Catalogue (ATCAL)\footnote{http://www.narrabri.atnf.csiro.au/calibrators/} and published LAS information, to confirm our sources are compact. Six sources had an LAS$>8''$ at 1.4 or 5\,GHz (using a 6\,km baseline) and were resolved at this frequency, and higher frequencies, despite the criterion imposed to have an angular size of $<10''$ from the AT20G.  
This suggests these sources do have extended emission not detected by AT20G, which primarily picks up core-dominated emission from AGN \citep{at20,at20gcat}. We note that these objects with extended radio emission may still be CSS or GPS objects, but have undergone a re-ignition of the radio AGN. The core of an AGN is detected at high frequencies ($>$20\,GHz), and if the AGN has restarted we would detect the core within an area of more diffuse radio emission on larger scales. \citet{hancock2} found several objects where it appears the AGN has restarted and is undergoing a second period of activity, and the previous generation of activity is detected at lower radio frequencies.

The six sources identified as extended from the ATCAL and other published data were removed from our list. The remainder of our objects had flux on scales $<8''$ at 1.4 and 5\,GHz and were considered to be compact, resulting in a list of 35 sources.

\begin{figure*}
\begin{center}
\includegraphics[scale=0.98]{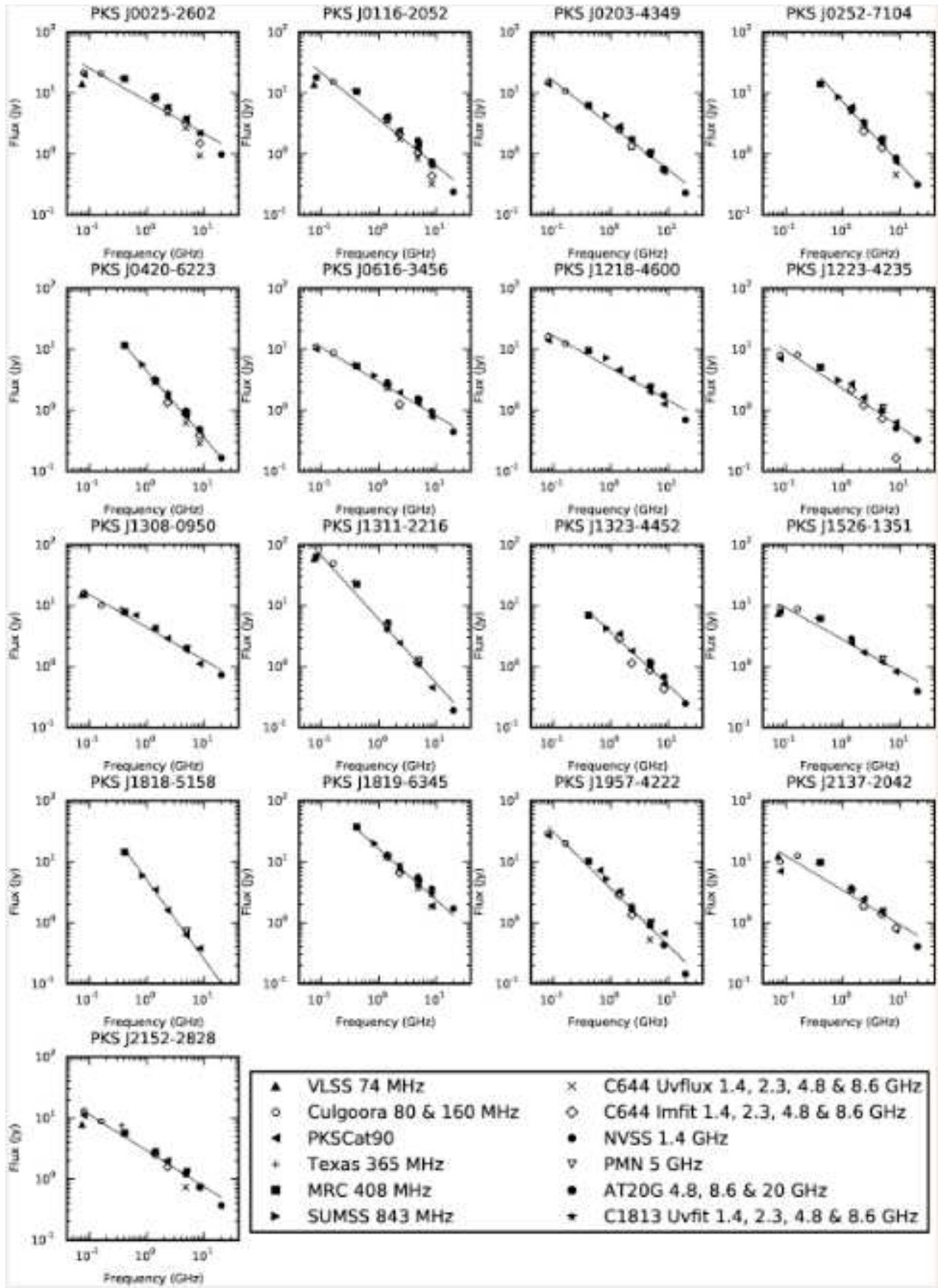}
\caption{Radio spectra for CSS sources from our sample, with power-law spectral fits overlaid. The frequencies are all in the observer's frame. The two new CSS candidates are PKS J1311-2216 and PKS J2152-2828. C644 and C1813 are the ATCA observing programs presented in \S~\ref{sec:obs}.}
\label{fig:css}
\end{center}
\end{figure*}

\begin{figure*}
\begin{center}
\includegraphics[scale=0.9]{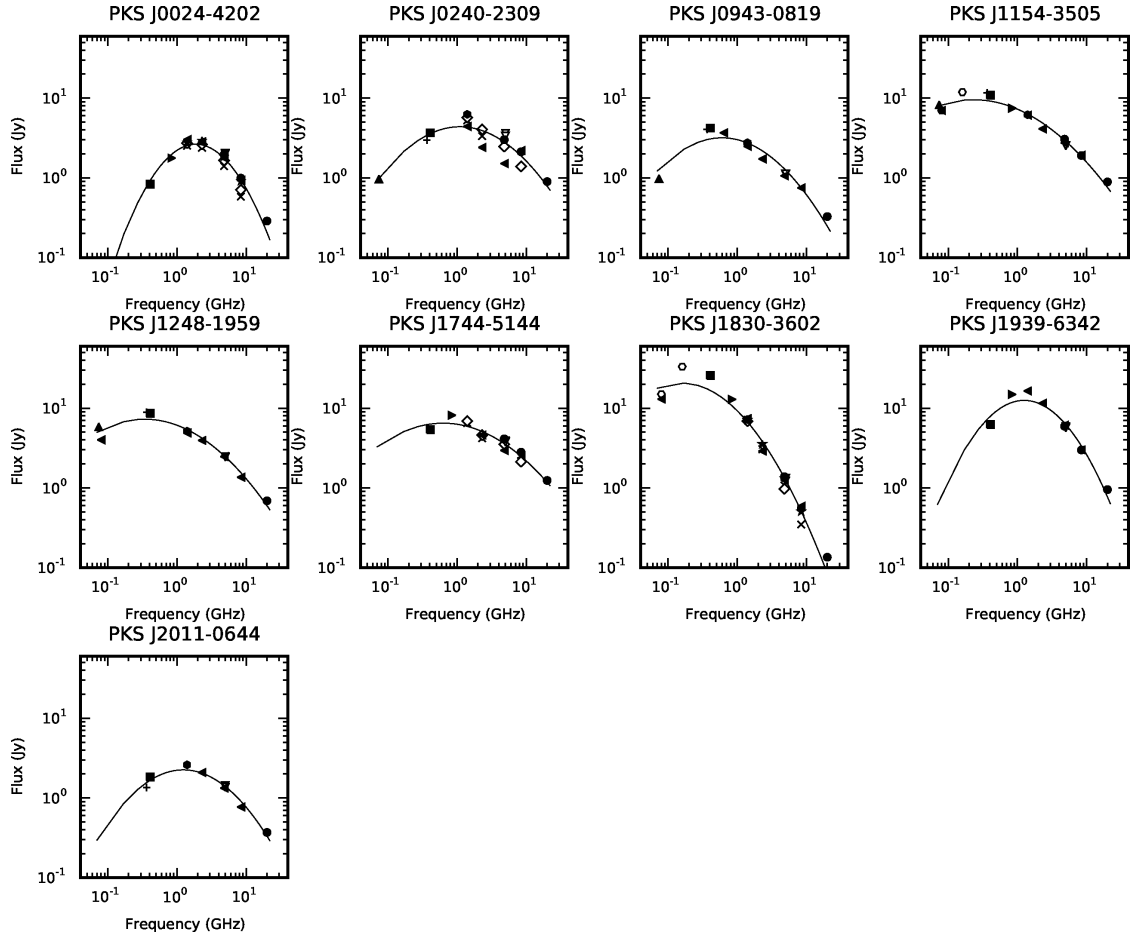}
\caption{Radio spectra for the GPS sources from our sample, with quadratic fits overlaid. The frequencies are all in the observer's frame.}
\label{fig:gps}
\end{center}
\end{figure*}

\begin{table}
\begin{center}
\caption{\textsc{Summary of Largest Angular Size Information for all sources}}
\label{table:las}
\begin{tabular}{|cccc|} \hline
Source & LAS & Frequency & Reference\\ 
& (arcsec) & (GHz) &\\ 
\hline
PKS J0024$-$4202&$<$1.8&9&1\\
PKS J0025$-$2602&$<$0.68&2.3&2\\
PKS J0116$-$2052&0.7&15&3\\
PKS J0203$-$4349&1.5&5&4\\
PKS J0240$-$2309&1.018&5&5\\
PKS J0252$-$7104&$<$0.24&2.3&2\\
PKS J0420$-$6223&$<$2.1&5&4\\
PKS J0616$-$3456&$<$1.8&5&4\\
PKS J0943$-$0819&0.05&5&5\\
PKS J1154$-$3505&$<$0.17&2.3&2\\
PKS J1218$-$4600&$<$1.9&5&4\\
PKS J1223$-$4235&$<$2.1&5&4\\
PKS J1248$-$1959&0.5&5&5\\
PKS J1308$-$0950&$<$0.46&2.3&2\\
PKS J1311$-$2216&1.1&5&1\\
PKS J1323$-$4452&$<$2&5&4\\
PKS J1526$-$1351&0.4&0.61&6\\
PKS J1744$-$5144&0.052&2.3&7\\
PKS J1818$-$5158&2.7&5&4\\
PKS J1819$-$6345&$<$0.41&2.3&2\\
PKS J1830$-$3602&$<$1.5&5&4\\
PKS J1939$-$6342&$<$0.07&2.3&2\\
PKS J1957$-$4222&2&5&4\\
PKS J2011$-$0644&0.03&5&5\\
PKS J2137$-$2042&0.25&2.3&2\\
PKS J2152$-$2828&$<$2&5&1\\
\hline
\end{tabular}\\
\textsc{Notes.}-- All frequencies are in the observer's frame.\\
\textsc{References.}-- (1) Stevens 2010 (private comm.), (2) Tzioumis et al. 2002, (3) Mantovani et al. 1994, (4) Burgess \& Hunstead 2006, (5) O'Dea 1998, (6) Mantovani, Muxlow \& Padrielli 1987, (6) Jauncey et al. 2003.  
\end{center}
\end{table}

\subsubsection{Selection Criterion 11: Physical Size}
\label{sec:physical}
To confirm that the radio emission resides entirely within the optical host galaxy, we applied a physical size cut of 20\,kpc as a final criterion. All except one of the 35 objects remaining had a published redshift, and this criterion ensured the removal of any sources with radio emission outside the host galaxy. The object without a redshift (the previously unknown GPS source) was removed at this point, as we cannot ensure it satisfies criterion 11. Out of these 34 sources, eight sources had a physical size scale$>20$\,kpc, and were thus removed from our sample. Our final sample consists of 26 sources, comprising 17 CSS candidates and sources and 9 GPS sources. LAS is shown as a function of redshift in Figure~\ref{fig:angsize}, together with the 15\,kpc and 20\,kpc linear size limits given by O'Dea, for the final 26 sources. The LAS data is presented in Table~\ref{table:las}.

\subsubsection{Comparison of Selection Criteria}
In Table~\ref{table:odea}, we show a summary of the selection criteria used by \citet{Review} compared to the selection criteria for our sample. In order to keep our selection criteria uniform, we imposed an angular size limit ($30''$ initially, and $8''$ for the final selection, see Table~\ref{table:selection}) rather than a physical size limit as used by O'Dea, as redshifts were not available for all 76 sources a priori. We also note that the O'Dea sample is selected in a more heterogeneous manner, from different catalogues available at the time of publication, for CSS and GPS sources separately (as noted in Table~\ref{table:odea}).

\subsubsection{Variability}
As the radio observations for each source span several decades in time, this provides an opportunity to identify variable sources. CSS and GPS sources generally are not intrinsically variable on short time scales \citep{aller,Review}. Table~\ref{table:var} shows the ten sources observed, and the flux densities, including the new observations (see \S\,~\ref{sec:obs}), compared to previous measurements, covering five epochs in time. Our flux density errors are $\sim10\%$ for all measurements in Table~\ref{table:var}. We found the flux densities were consistent within 15\% at 1.4 and 2.3\,GHz and 8\% for 4.8 and 8.4\,GHz for all sources except one. The source with higher variability ($>15\%$ at all frequencies) was already catalogued as a CSS source, as discussed in \S~\ref{sec:coolsources}. This variability may be an artifact arising from missing extended flux with the different surveys, or could potentially be genuine variability over the several decades spanned. 

For the rest of the sources in our sample catalogued as CSS/GPS, the archival data samples fewer epochs. This existing data suggests the remaining sources have flux density variations within 10\% for the measured frequencies. We conclude that none of our sources are significantly variable.  

\subsubsection{The final sample}
In summary, the 76 sources from the initial selection fall into four categories: 
\begin{itemize}
\item 2 new CSS candidate sources,
\item 15 known CSS sources (Figure~\ref{fig:css}),
\item 50 poor CSS candidates and/or known sources, with data broadly spread around the power-law fit, an LAS $>10''$ or a double or tripled lobe radio morphology (spectra not shown), and
\item 9 known GPS sources, distinguished by a turnover in their spectrum (Figure~\ref{fig:gps}).
\end{itemize}
The catalogue for the final 26 sources is presented in Section~\ref{sec:results}.
\normalfont

\section{Observations}
\label{sec:obs}
Of the 76 sources remaining following the selection using the initial criteria of \S~\ref{sec:crit}, 55 were observed with the ATCA, and details of the observations are summarized in Table~\ref{table:atca}. The new observations presented in this paper are shown in Table~\ref{table:c1813}.

\subsection{ATCA Project C644}
\label{sec:c644}
ATCA observations for project C644 (PI King) were made in July 1997, over 45.5 hours, using the 6A Array, full 128\,MHz bandwidth, all polarization products, at $\nu_{o}=$1.384, 2.496, 4.8 and 
8.64\,GHz. Sources were observed in snapshots of one minute three to four times each, since only flux density information, rather than high quality images, were desired. 
Only a primary amplitude calibrator was used, as the sources are bright enough to self-calibrate. 38 sources from our initial sample were observed, including 10 from the final sample. Flux 
densities were measured both from the visibility data and also from images using the MIRIAD tasks \textsc{Uvflux} and \textsc{Imfit} respectively. The 
images were made with the standard tasks \textsc{Invert}, \textsc{Clean} and \textsc{Restor}, with an average rms of 0.05\,Jy. For 30\% of these observations the ($u, v$) coverage from the snapshots was insufficient to measure a reliable flux density.

\subsection{ATCA Project C1813}
\label{sec:c1813}
For project C1813 (PI Randall) 52 sources were observed on 19-20 June 2008. Observations were made in the 
1.5B array, using 128\,MHz bandwidth for each of four frequencies centered on $\nu_{o}=$1.384, 2.496, 4.8 and 8.64\,GHz. Two observations of four minutes each were made on each source at each frequency, and the primary calibrator PKS B1934-638 was observed for ten minutes at the beginning of each observing run. No secondary 
calibrators were used, as all sources are bright enough to self-calibrate.

Flux density information was measured from the ($u, v$) data for each source, and added to the catalogued data for 10 sources in the final sample. Data for the other 42 observed sources is presented in Table~\ref{table:c1813} (an asterisk indicates those which were not among the 76 initial CSS candidates), whilst the data for the sources from our final sample are included in Table~\ref{table:sample}. The typical errors for these measurements are $\sim10\%$, as described in \citet{Middelberg08}, given that we take systematic errors of the telescope systems into account, incorporating errors in the pointing and flux calibration.

\begin{table*}
\begin{center}
\caption{\textsc{Sources measured at multiple epochs}
\label{table:var}}
\begin{tabular}{|cccccccccccccc|} \hline
Source & S$_{1.4}$ & S$_{1.4}$ &S$_{1.4}$ &S$_{2.3}$&S$_{2.7}$ &S$_{4.8}$ &S$_{4.8}$ &S$_{4.8}$&S$_{4.8}$ &S$_{8.6}$&S$_{8.6}$ &S$_{8.6}$\\
 & C1813 & NVSS & PKSCAT90 & C1813 & PKSCAT90 & C1813 & PMN & PKSCAT90 & AT20G & C1813 & PKSCAT90 & AT20G\\
 & Jy & Jy & Jy & Jy & Jy & Jy & Jy & Jy & Jy & Jy & Jy & Jy\\
\hline
PKS J0024$-$4202&2.78&...&3.02&2.93&2.84&1.85&2.04&1.77&1.96&0.91&0.93&0.99\\
PKS J0025$-$2602&8.37&8.75&8.6&5.94&5.8&3.64&3.75&3.76&...&2.15&2.2&...\\
PKS J0116$-$2052&3.81&4.091&4.1&2.44&2.4&1.27&1.42&1.24&1.66&0.64&0.67&0.75\\
PKS J0203$-$4349&2.67&...&2.87&1.81&1.7&1.03&1.06&1.01&1.025&0.58&0.53&0.57\\
PKS J0252$-$7104&5.59&...&5.9&3.42&3.1&1.69&1.72&1.54&1.79&0.80&0.74&0.86\\
PKS J0420$-$6223&3.26&...&3&1.94&1.68&0.93&0.92&0.84&...&0.47&0.39&...\\
PKS J1819$-$6345&13.34&...&12.3&8.80&7.4&5.14&4.51&4.21&...&2.95&1.89&...\\
PKS J1830$-$3602&7.12&7.24&7.4&3.54&2.9&1.32&1.32&1.27&1.38&0.51&0.59&0.55\\
PKS J1957$-$4222&3.25&...&3.27&1.90&1.6&0.92&1.01&0.89&0.91&0.43&0.67&0.43\\
PKS J2152$-$2828&2.75&2.87&2.8&1.98&2&1.25&1.32&1.32&1.20&0.75&0.75&0.73\\
\hline
\end{tabular}\\
\textsc{Notes.}-- Flux densities given in this table cover five different epochs of time, and all errors on flux density measurements are $\sim10\%$. Frequencies are all in the observer's frame. All PKSCAT90 measurements were taken before 1990, and could be from as early as 1965. C1813 observations were completed in 2008 and AT20G data measurements were completed between 2004 and 2007 (data spanning multiple frequencies was simultaneously taken). NVSS measurements were taken between 1993 and 1996, and PMN was completed in one year, in 1990. \\
\end{center}
\end{table*}

\section{Results and Interesting Sources}
\label{sec:results}
\subsection{Spectral Classification and Source Catalogue}
\label{sec:class} 
Our resulting catalogue consists of 2 new CSS candidate sources, 15 known CSS sources and 9 known GPS sources. Although our sample is not complete (due to initial criterion 1 ensuring each source had at least four flux density measurements in PKSCAT90), we have formed an unbiased sample of these objects. We define a complete sample to be a set of objects from a parent population that includes all such objects that satisfy a set of well-defined selection criteria, and an unbiased sample to be one that is a subset of a complete sample, where the selection of the objects does not depend on the intrinsic properties of the sources. Our sample is unbiased and we estimate it is $\sim36\%$ complete, based on the number of sources excluded by the criterion requiring four flux densities. With future work, we can expand the sample to ensure it is complete.

Radio spectra for the final 26 sources are shown in Figures~\ref{fig:css} and \ref{fig:gps}. The radio spectra show all radio data points compiled from the various catalogues and surveys (Table~\ref{table:cats}). For  our CSS candidates and sources, the fitted least squares power law of the form $S\propto\,\nu^{\alpha_{fit}}$, is shown. The quadratic fit performed for the candidate selection (\S~\ref{sec:secondary}) is shown on the GPS sources. 

Our CSS candidates and sources follow a strong power-law spectrum across two orders of magnitude in frequency, with few deviations. Any deviations or outlying points from the power-law fits may be attributed to, (a) measurement errors, (b) limitations of the data itself, such as minimal sampling of the ($u,v$)-plane, or (c) genuine deviations from the power law, perhaps caused by electron cooling.

All 26 remaining sources are listed in Table~\ref{table:sample}, with a `k' tag in the name column if the source was previously known to be a CSS or GPS object, and a `C' or `G' for CSS or GPS candidate respectively. A selection of the sample properties, source name, classification, right 
ascension and declination, redshift, R-band magnitude, optical classification, physical size, overall radio spectral index $\alpha_{fit}$, selection spectral index $\alpha^{5}_{2.7}$ and several radio flux densities are given. Also, where a spectral index $\alpha_{fit}$ is listed for GPS sources, this spectral index is taken from a power law fit to the part of the radio spectrum at frequencies above the spectral peak. The online version of this table includes all radio flux density measurements for all sources, and all relevant optical data. We also present the LAS data in Table~\ref{table:las}, and Figure~\ref{fig:angsize}.
\subsection{Optical Properties of CSS and GPS sources}
\label{sec:props}
\subsubsection{Optical Counterparts and Morphology}
Optical counterparts for 22 of the 26 sources have been determined from the SuperCOSMOS Science Archive (based up digital scans of the 1.2\,m UK Schmidt Telescope and Palomar-II Oschin Schmidt Telescope sky surveys in the red, blue and near-infrared, taken between 1974and 2002). The optical counterparts are determined for each radio source by selecting the optical source closest to the radio position from the AT20G survey \citep{at20,at20gcat} within a radius of $6''$. We used NED to identify whether we had any interacting galaxies or galaxy pairs. We further use the optical morphological classification from SuperCOSMOS, where an object is classified as a `galaxy', `star' (or QSO) or unclassifiable. The optical morphology classification for all objects consisted of 18 galaxies and 8 star-like objects or unresolved objects, likely QSOs. For the four sources too faint to be detected with SuperCOSMOS we use the position, classification, $B_{J}$ magnitude and R band and I band magnitudes or limits available from the literature for our analysis and in Table~\ref{table:sample}: \citet{burgess} for PKS J0420$-$6223 and PKS J1744$-$5144, \citet{snell} for PKS J0116$-$2052 and \citet{mcc} for PKS J2011$-$0644 for our analysis and in Table~\ref{table:sample}. The Burgess \& Hunstead $B_{J}$ band magnitude is from COSMOS, an earlier version of SuperCOSMOS \citep{opt}. 

It has been suggested \citep{Review,cfanti,holtb} that a significant fraction of CSS and GPS hosts are interacting. Of the 64 sources in the O'Dea sample, 57\% of the GPS sources show a disturbed nucleus and 13\% have a second nucleus nearby. Similarly, \citet{geld,geld1} suggests all of his sample of 20 CSS sources show evidence of interaction or disturbance. \citet{heck} found that 54\% of their powerful radio galaxies had distorted nuclei, and 20\% had a close companion. \citet{Review} suggests similar statistics for sample of CSS sources.

In our sample we see seven sources interacting and/or distorted with close companion(s), two distorted sources, one possible double nucleus and 16 isolated sources. This is 27\%, 8\%, 4\% and 61\% of our sample respectively. We are thus seeing a much lower level of interaction or disturbance, in this sample (up to $\sim\,39\%$) compared to much higher fractions in those previously published. Further deeper optical imaging and spectroscopy would reveal whether these objects are all from the same parent population, by removing resolution limitations and allowing us to determine whether nearby objects in the images are associated or merely appearing in projection. This morphological difference between our sample and the O'Dea sample suggests that the host galaxy properties of the CSS and GPS population may not be as well understood as previously suggested. Three optically unusual systems in our sample, PKS J1819$-$6345, PKS J0616$-$3456 and PKS J1223$-$4235, are further discussed, in Section~\ref{sec:coolsources}. They consist of a blue disk host galaxy, a galaxy pair and a pair of merging galaxies.

\begin{figure}
\begin{center}
\includegraphics[scale=0.4]{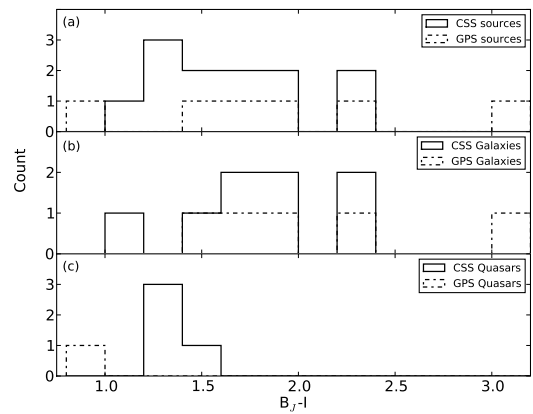}
\caption{Panel (a) is the SuperCOSMOS B$_J$-I distribution for all 26 sources with optically identified counterparts, split into CSS and GPS sources. Panel (b) shows the distribution for CSS and GPS galaxies, while Panel (c) shows the CSS and GPS quasars.}
\label{fig:opt2}
\end{center}
\end{figure}

Figure~\ref{fig:opt2} shows the SuperCOSMOS $B_{J}-I$ distribution for our sample, split into (a) CSS and GPS sources, (b) CSS and GPS galaxies, and (c) CSS and GPS quasars. Figure~\ref{fig:morph} shows all 26 of the SuperCOSMOS $B_{J}$ images in  from our sample. We present $B_{J}$ images here as this is the optical band most likely to highlight morphological distortions associated with star formation, although in our classification we used multiple bands ($B_{J}$, R and I) to maximize our chances of identifying distorted or interacting sources.

\subsubsection{Redshift and Populations of sources}
\begin{figure*}
\begin{center} 
\includegraphics[scale=0.4]{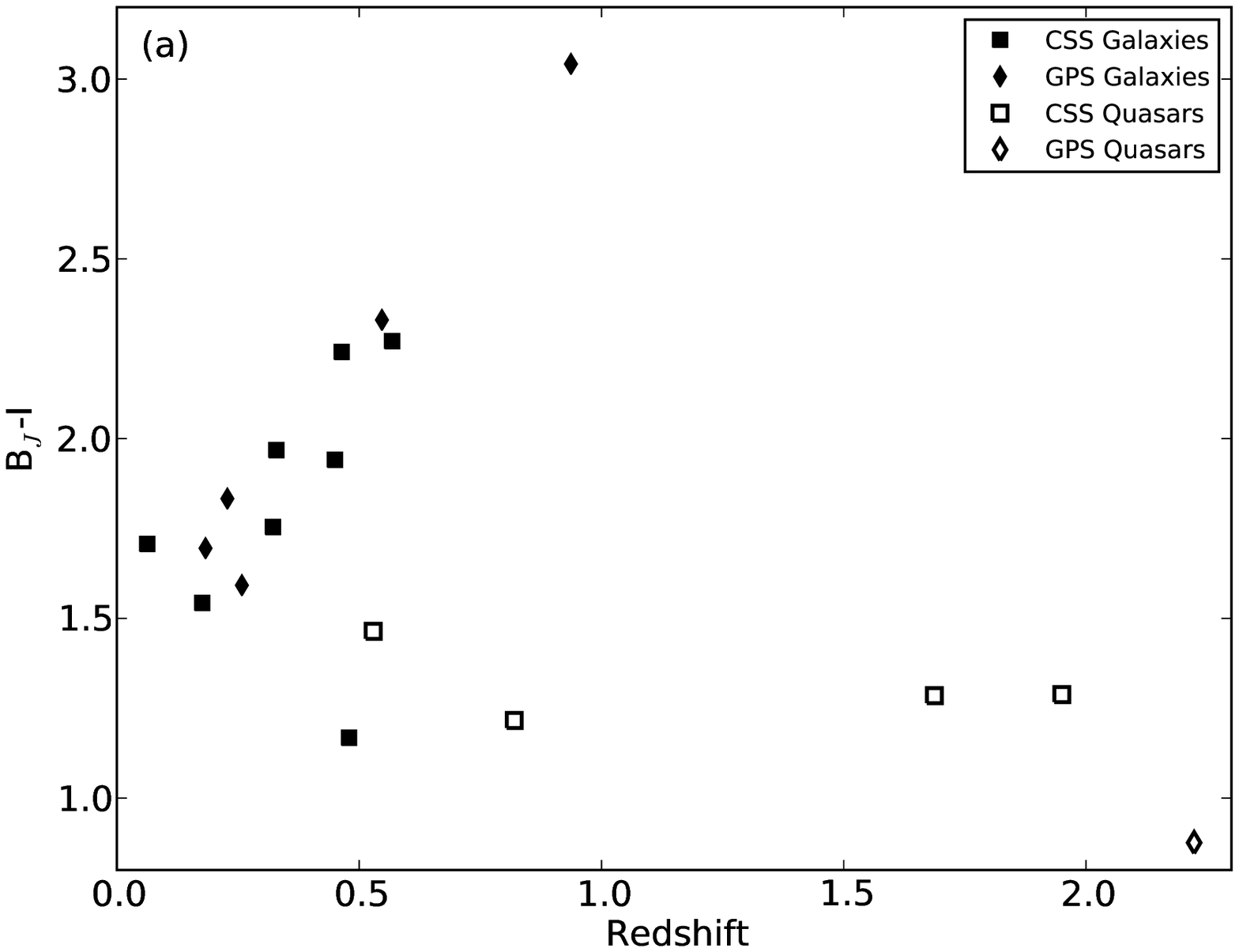}\includegraphics[scale=0.4]{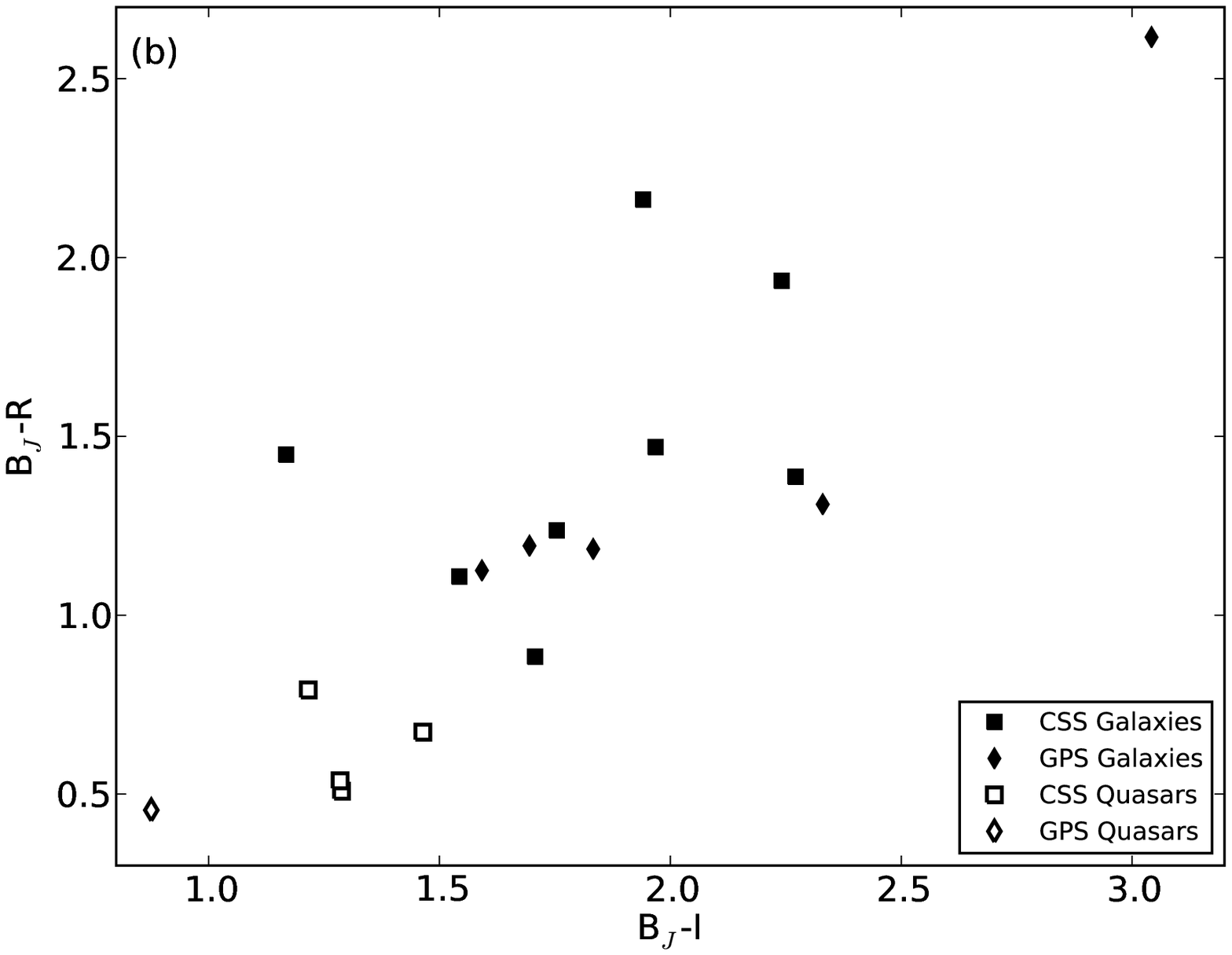}
\caption{(a) SuperCOSMOS B$_{J}-I$ magnitudes against redshift for our sample only. It suggests that we are seeing two different host galaxy types, LRGs and quasars. (b) SuperCOSMOS B$_{J}-I$ magnitudes plotted against B$_{J}-R$ for our sample. The symbols are the same as for the left figure. The very blue colours of the QSO sources, and the red colours of the galaxies, emphasise the likelihood of two populations of host galaxy within out sample.}
\label{fig:opt}
\end{center}
\end{figure*}

Redshifts were obtained for all sources using NED, including 6 photometric and 20 spectroscopic redshifts. In Figure~\ref{fig:opt}, we show redshift versus $B_{J}-I$ and $B_{J}-I$ versus $B_{J}-R$. These distributions suggest that we see two populations of objects, quasar-like sources, with blue colours, higher optical luminosities, and extending to higher redshifts;  and luminous red galaxy (LRG)-type objects with redder colours, and restricted to lower redshifts. In Figure~\ref{fig:opt}(a), we see the quasars as the blue objects with higher redshifts along the bottom, and the LRGs as the redder, nearby objects. In Figure~\ref{fig:opt}(b), the population split between LRGs and quasars can still be seen, and we see the quasar like objects clustering to bluer colours in the bottom left, with the red LRG type objects toward the top right. It is also clear that the split between quasar-like objects and LRGs is a gradual change, as we would otherwise see a bimodal distribution in the $B_{J}-I$ colours. We note that the error in SuperCOSMOS magnitudes is typically $0.3-0.5$ magnitudes, and we treat these distributions with some caution. In addition, there is a 15 year gap between the $B_{J}$ and R magnitudes used in SuperCOSMOS, and if our sources are optically variable, this may also have an effect. 

\begin{figure*}
\begin{center}
\includegraphics[scale=0.4]{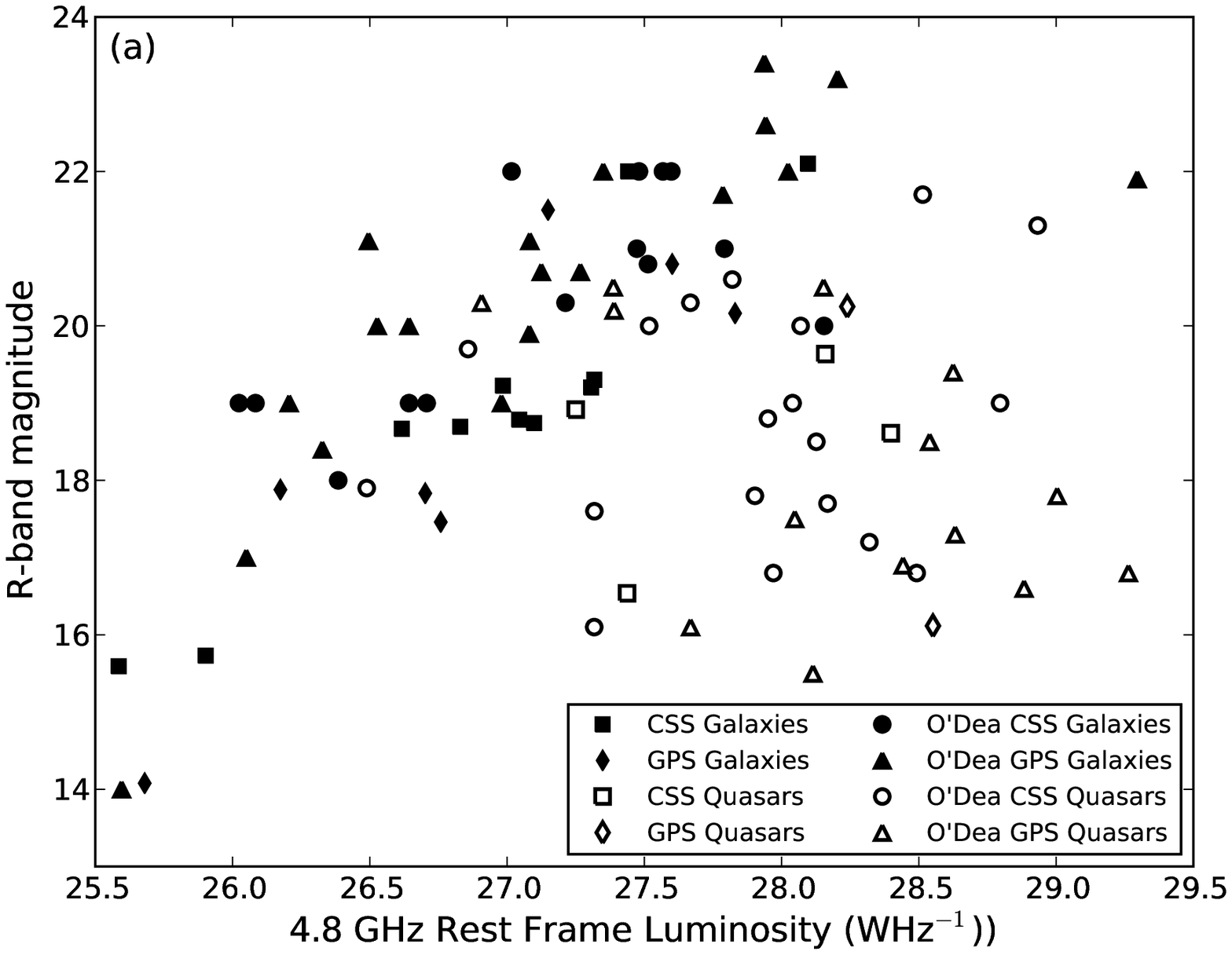}\includegraphics[scale=0.4]{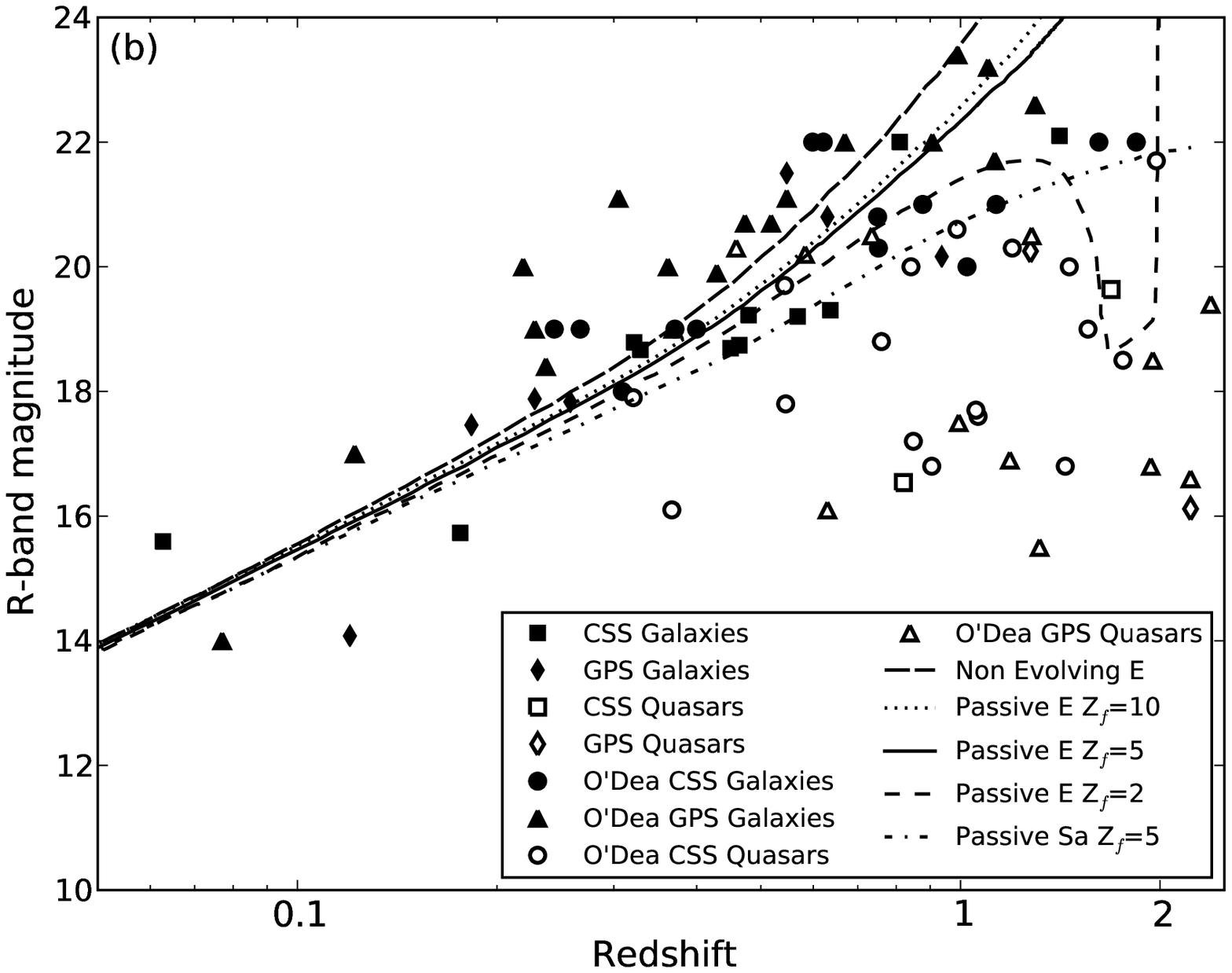}
\caption{(a) SuperCOSMOS R-band magnitudes against 4.8\,GHz rest-frame luminosities for our sample and the O'Dea sample. This illustrates a relatively tight correlation between optical magnitude and radio luminosity for LRG-type systems, which does not hold for the quasar population, where the optical emission is likely to be dominated by that from the AGN. (b) SuperCOSMOS R-band magnitudes against redshift for both samples. The models are from \citet{bruz} and are represented as follows: \textit{long-dashed curve}, non-evolving elliptical (E) galaxy,  \textit{dotted curve}, E galaxy with $z_{f}=10$, \textit{solid curve}, E galaxy with $z_{f}=5$, \textit{short-dashed curve}, E galaxy with $z_{f}=2$ and \textit{dashed-dotted curve}, Sa galaxy with $z_{f}=5$.}
\label{fig:opt3}
\end{center}
\end{figure*}

Figure~\ref{fig:opt3}(a) and (b) shows R-band magnitude versus both rest-frame luminosity and redshift, for both our sample and the O'Dea sample. The CSS and GPS sources are also split here by optical classification as a galaxy or quasar. Overall, it appears that we are looking at similar populations. A luminosity cut does not affect the overall appearance, or remove any one area of the sample. The models are from \cite{bruz} and model host galaxy stellar populations rather than AGN properties. The fact that these models closely track the LRG population emphasises that the bright magnitudes of the quasar population are dominated by optical emission from the AGN rather than stellar population.
\begin{figure*}
\begin{center}
\includegraphics[scale=0.76]{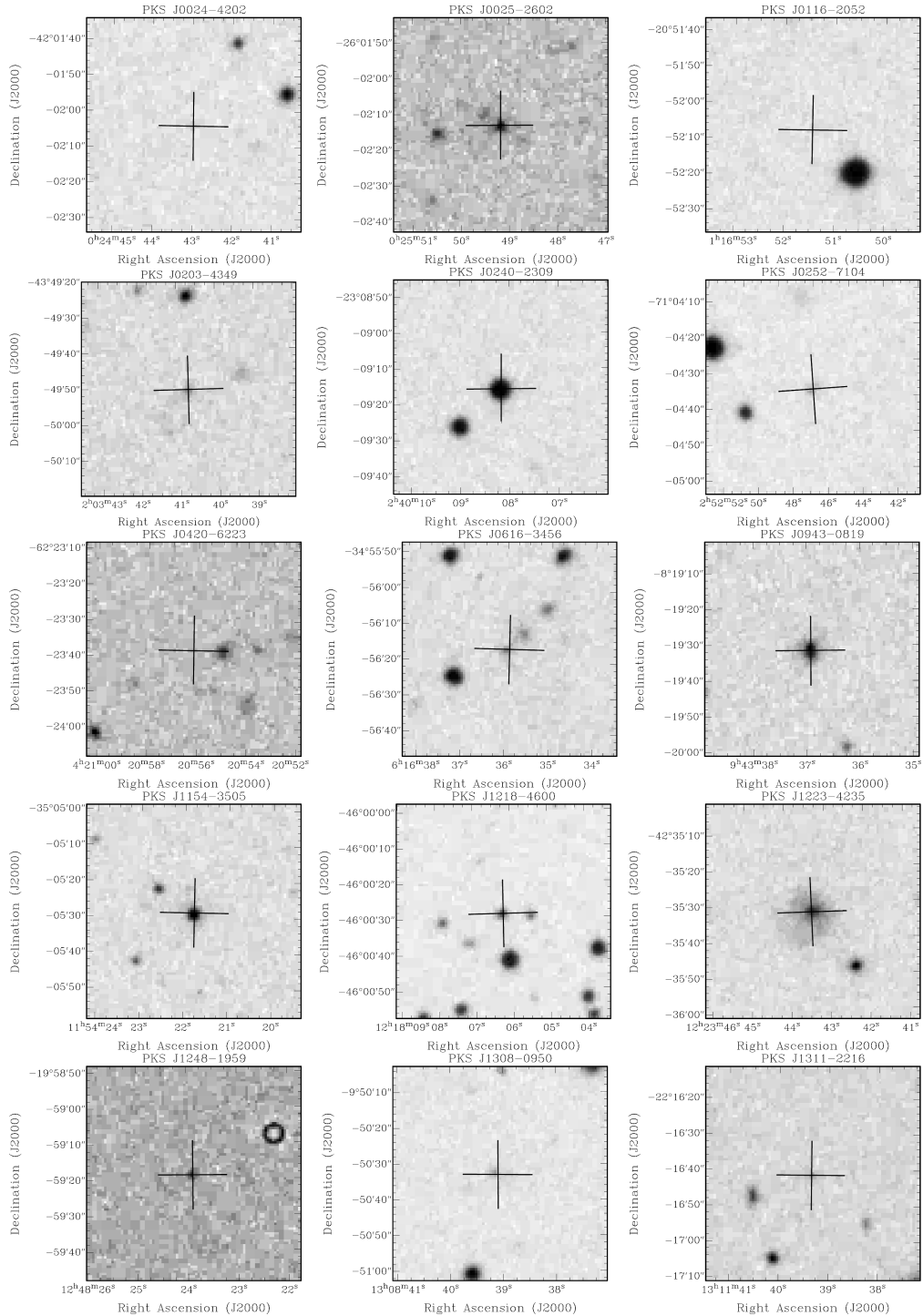}
\caption{SuperCOSMOS B$_J$ images for our entire sample, with a cross indicating the optical counterpart position. Images are 1$'$ square. PKS J0420$-$6223 and PKS J1744$-$5144 have a cross indicating the optical counterpart position from Burgess \& Hunstead (2006), PKS J2011$-$0644 has a cross indicating the optical counterpart position from \citet{snellen} and PKS J0116$-$2052 has a cross indicating its optical position from \citet{mcc}. Several sources (such as PKS J1818$-$5158) appear to only just be detected by SuperCOSMOS and as such have very faint counterparts.}
\label{fig:morph}
\end{center}
\end{figure*}

\begin{figure*}
\begin{center}
\includegraphics[scale=0.92]{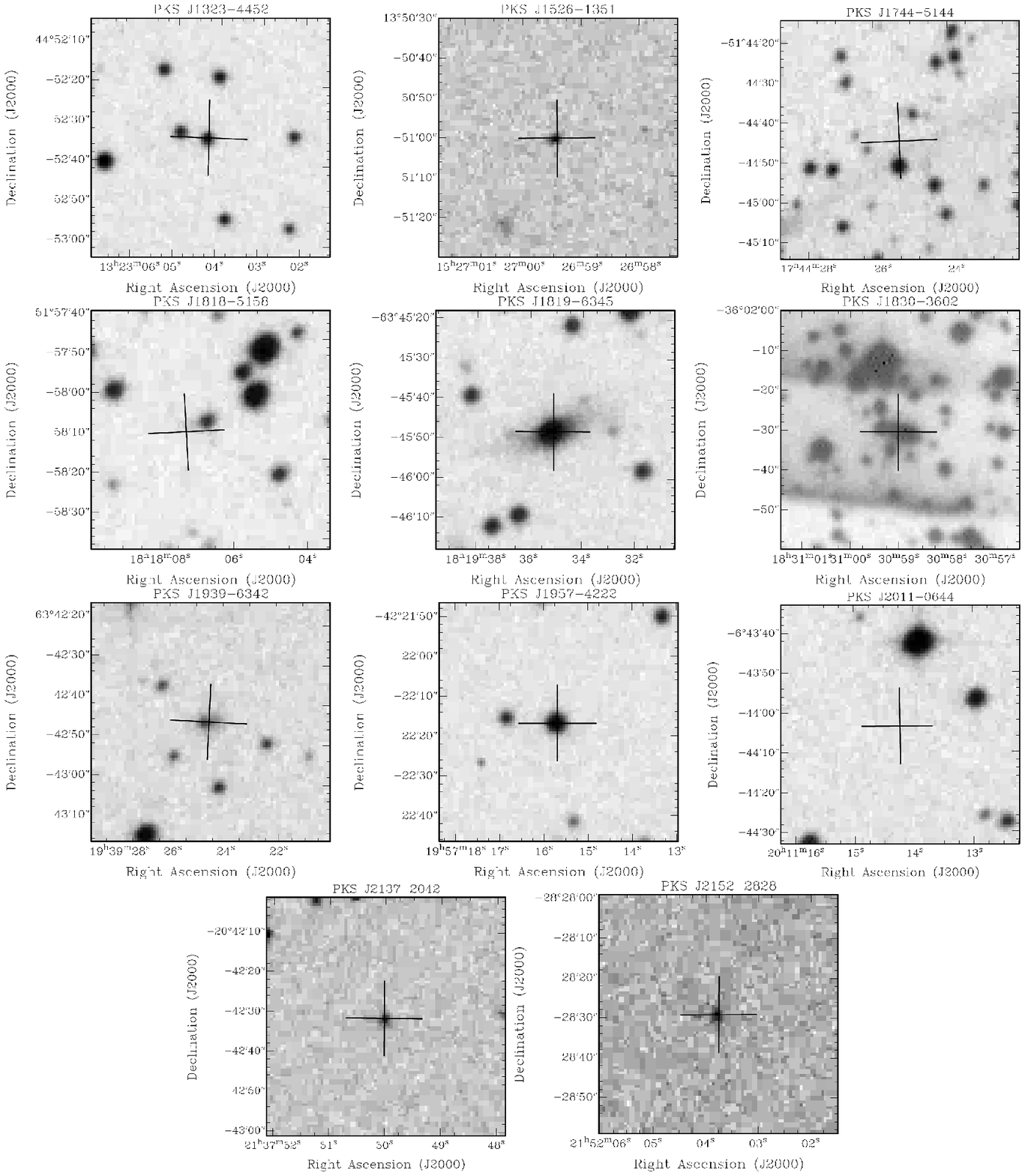}
\contcaption{}
\label{fig:morph1}
\end{center}
\end{figure*}

While these results also indicate that we are probing similar redshift, optical colour and luminosity regimes, our optical host galaxies are not predominantly disturbed or interacting as the O'Dea sample are. It remains unclear what the origin of this discrepancy may be. The difference in resolution of the optical imaging used by O'Dea and SuperCOSMOS is small. The resolution of the O'Dea imaging is only $\sim$1-2 times better than SuperCOSMOS for a few sources, and we would still be able to clearly see any double nuclei or companions in the SuperCOSMOS images. It is also important to note that classifying a source as distorted can be somewhat subjective, unless relying on more quantitative measures, such as the asymmetry index in the CAS scheme \citep{cas}.

\subsection{Three optically unusual CSS sources}
\label{sec:coolsources}
\begin{figure}
\begin{center} 
\includegraphics[scale=0.38]{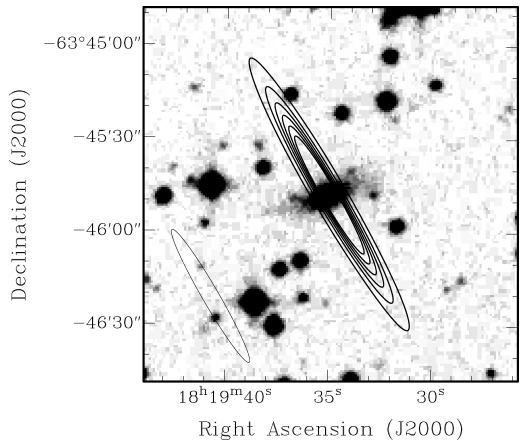}
\caption{B$_{J}$ band SuperCOSMOS image of PKS J1819-6345, with the  $\nu_{o}=$4.8\,GHz radio contours and synthesized beam size (lower left) from ATCA overlaid. The radio contours begin at 10\% of the peak flux (5.05\,Jy) and increase in increments of 15\%. This is an example of bright point source radio emission in a blue (B$_{J}-I=1$) disk galaxy, possibly hosting ongoing star formation.
\label{fig:j1819}}
\end{center}
\end{figure}

PKS J1819$-$6345 is catalogued as a CSS source by \citet{holta} and has an optical counterpart (Figure~\ref{fig:j1819}) which is a blue disk galaxy (B$_J-I=1.0$) at a redshift $z=0.063$. This object is unusual compared to the rest of our sample, with an optical host galaxy showing a blue disk. The blue colour of the galaxy may be an indicator of recent or ongoing star formation, possibly triggered by the AGN. The source has rest-frame L$_{ \nu_{r}=4.8}=10^{25.6}$\,WHz$^{-1}$, and it has a spectral index of $\alpha_{fit}=-0.86$. This source is also quite variable across the decades between radio observations (See Table~\ref{table:var}), and the variability we see may be due in part to star formation. 

Another optically unusual source is CSS object PKS J0616$-$3456, a galaxy pair. The galaxies are at a redshift of $z=0.329$, with B$_J-I=2.48$ and B$_J-I=2.88$. Although this is optically a galaxy pair, the radio position is centred on the first optical counterpart, and due to the compact nature of these radio objects, it is likely the radio object is only associated with one of the optical pair. The source has rest-frame L$_{ \nu_{r}=4.8}=10^{26.6}$\,WHz$^{-1}$, and it has a spectral index of $\alpha_{fit}=-0.56$. 

The final unusual source, PKS J1223$-$4235, is a well studied interacting pair of galaxies, where the host galaxy of the radio source appears to be merging with a much smaller galaxy. This object is interesting for several reasons; first, it is one of the closest CSS sources, at a redshift of $z=0.176$. Second, it has been studied intensively in the optical and radio \citep{safouris,johnston} and has been shown to host three different aged stellar populations, one of which is thought to be triggered by the merging of the small companion galaxy with the larger galaxy hosting the radio source. This object has an overall B$_J-I=1.5$, with a rest-frame L$_{ \nu_{r}=4.8}=10^{25.9}$\,WHz$^{-1}$, and it has a spectral index of $\alpha_{fit}=-0.63$.

\section{Discussion}
\label{sec:discussion}
\subsection{Comparison with previous sample}
\label{sec:comparison}
We compare our redshift, spectral index $\alpha_{fit}$ and luminosity distributions with the sample from \citet{Review}, which consists of 34 CSS and 33 GPS sources. Our sample was constructed to investigate the same parent population as the O'Dea sample, as it is the most intensively studied sample of CSS and GPS sources to date, but nonetheless our sample differs slightly in several ways. We note that due to the areas covered by the surveys from which our sample is drawn, there is some small overlap with the selection of the O'Dea sample. 

Four of the O'Dea sources are within our sample, and an additional five of the O'Dea sample in the southern hemisphere do not satisfy all of our initial selection criteria (\S\,\ref{sec:crit}).This is further discussed in \S\,\ref{sec:evolution}. Table~\ref{table:odea} shows the differences between the O'Dea sample selection and ours.

\begin{figure*}
\begin{center}
\includegraphics[scale=0.4]{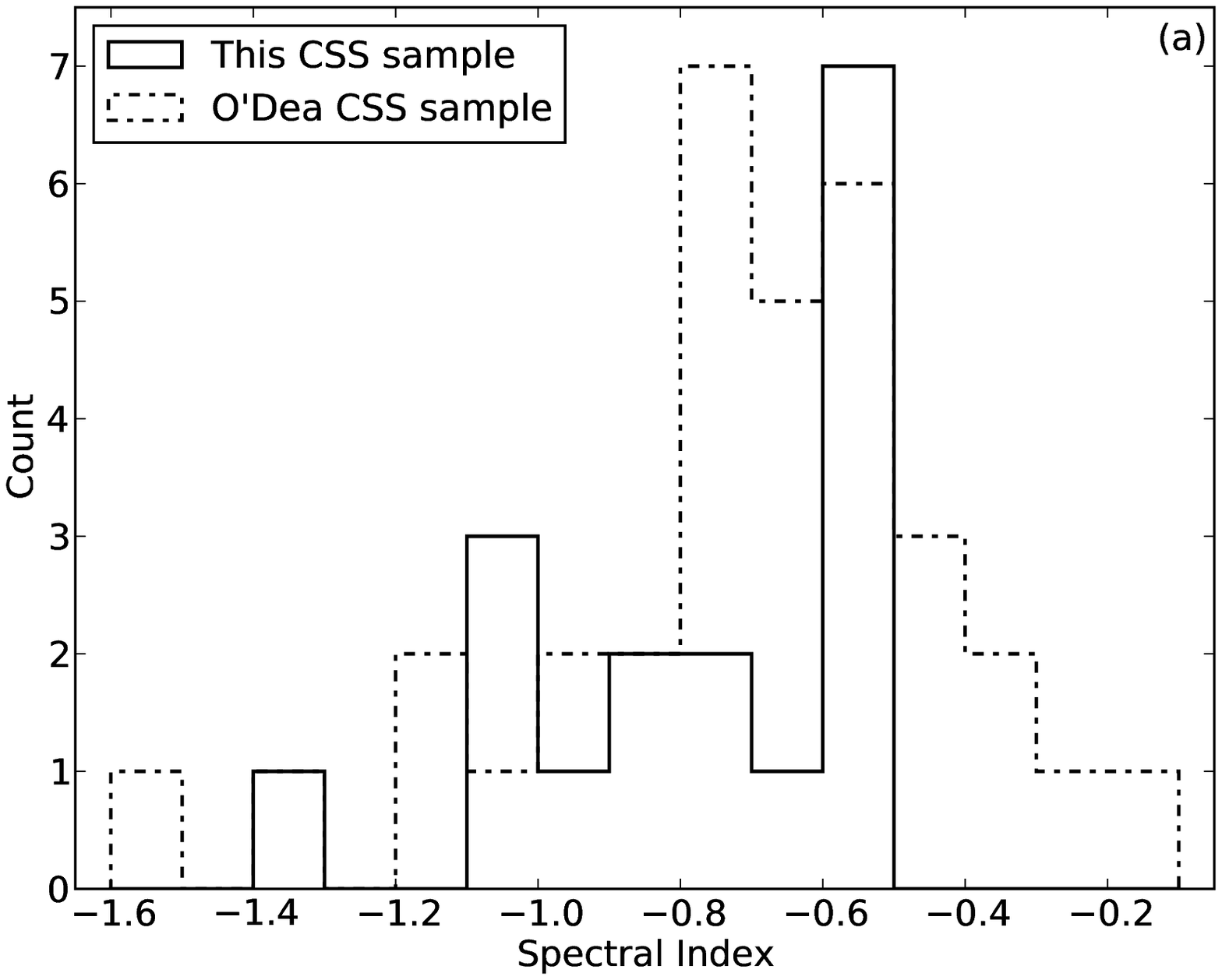} \includegraphics[scale=0.4]{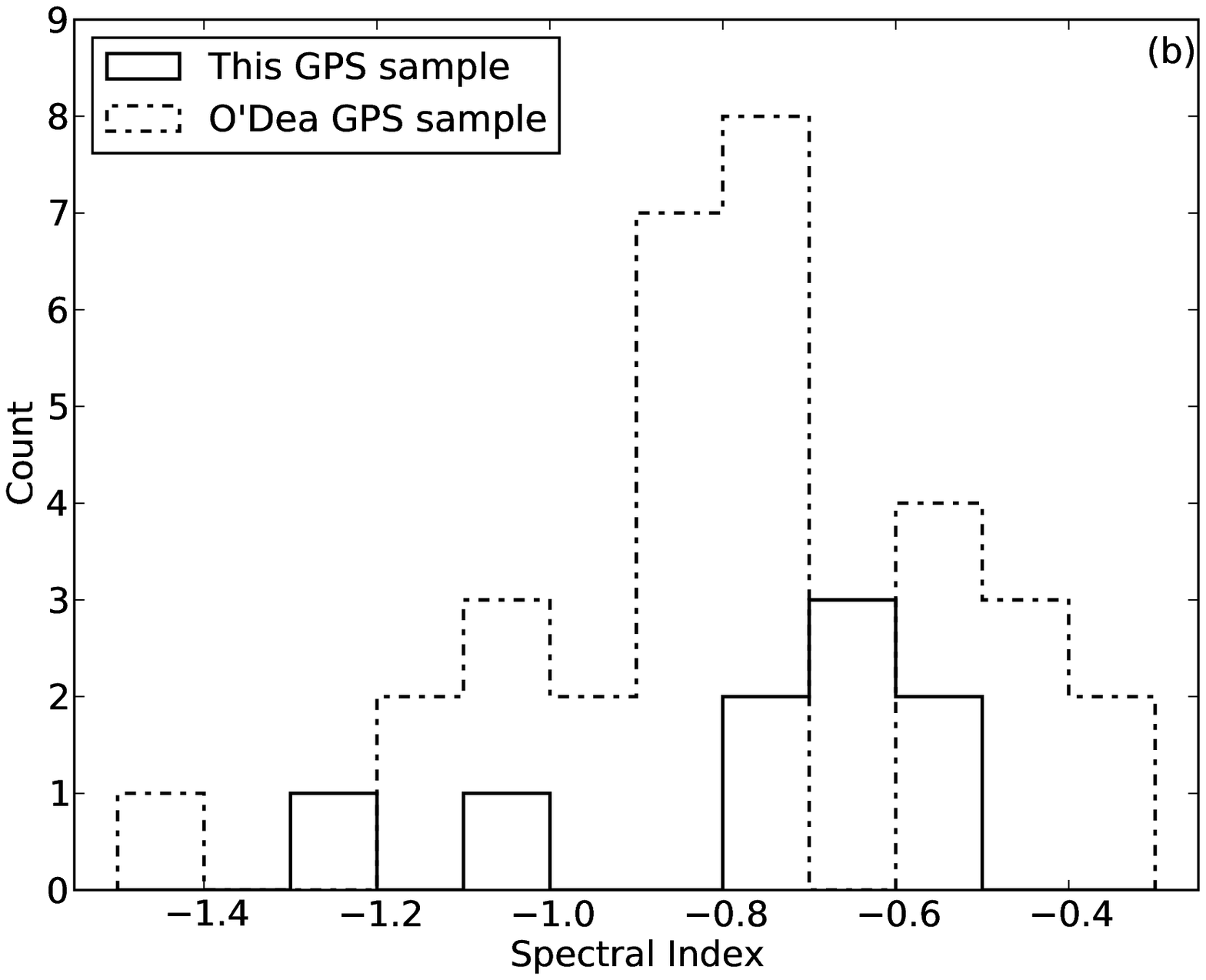}
\caption{(a) Spectral Index $\alpha_{fit}$ for this sample and the O'Dea sample for CSS sources, (b) Spectral Index $\alpha_{fit}$ above the peak for this sample and the O'Dea sample for GPS sources.}
\label{fig:specind}
\end{center}
\end{figure*}

\begin{figure*}
\begin{center}
\includegraphics[scale=0.42]{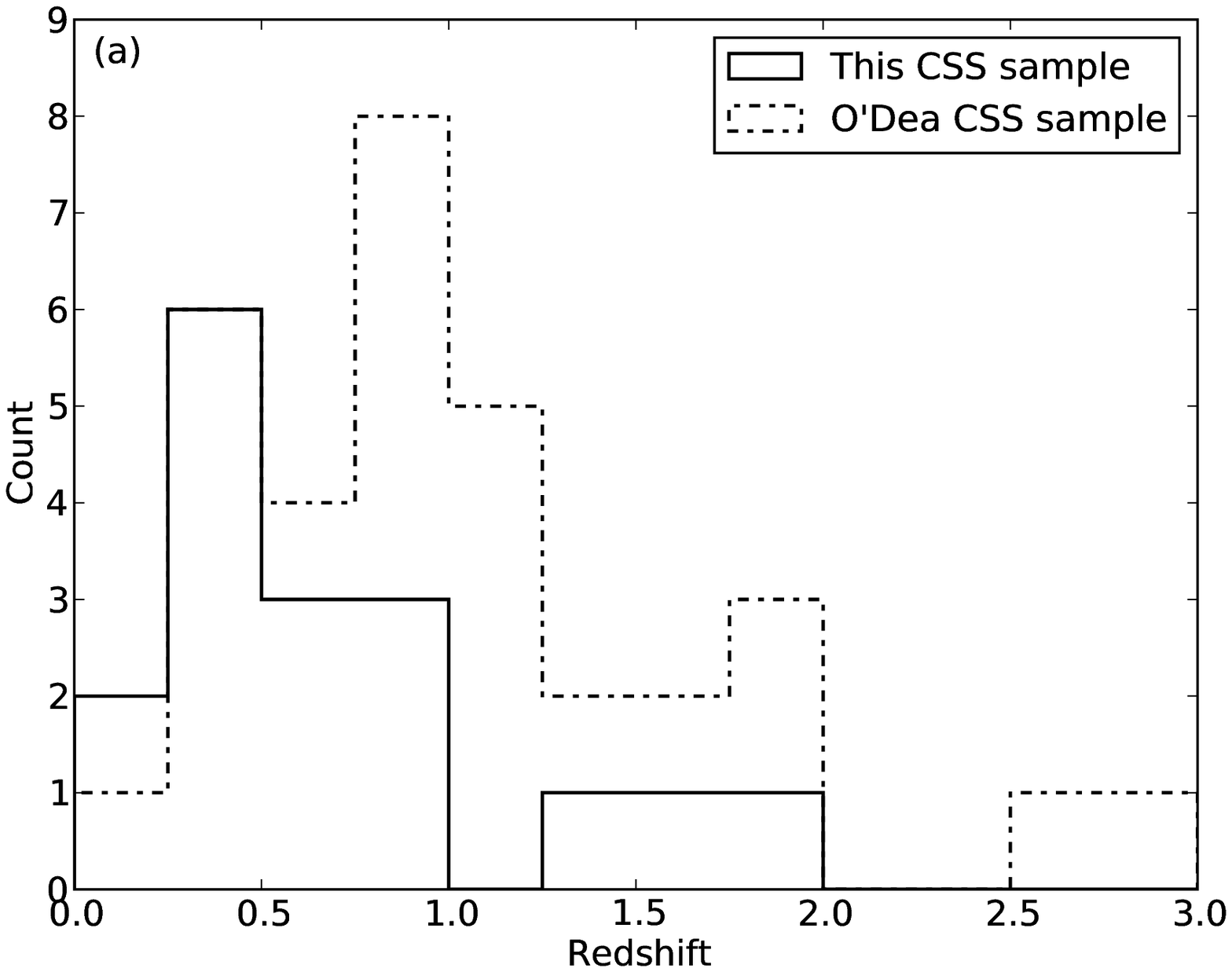} \includegraphics[scale=0.42]{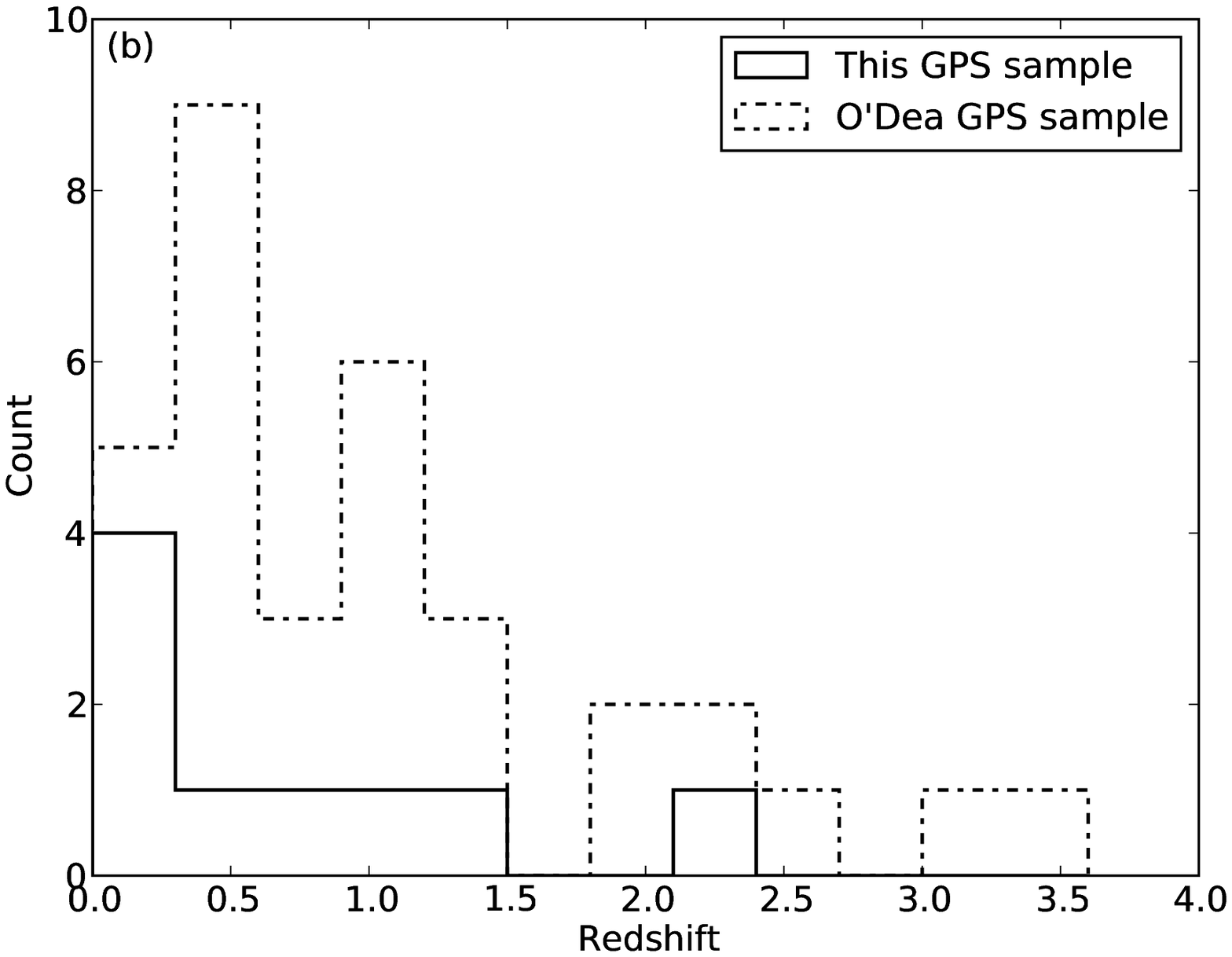}
\caption{(a) Distribution of redshifts for this sample and the O'Dea sample of CSS sources, (b) Distribution of redshifts for the GPS sources. Our sample generally spans lower redshifts than the O'Dea sample.}
\label{fig:redshift}
\end{center}
\end{figure*}

\begin{figure*} 
\begin{center} 
\includegraphics[scale=0.4]{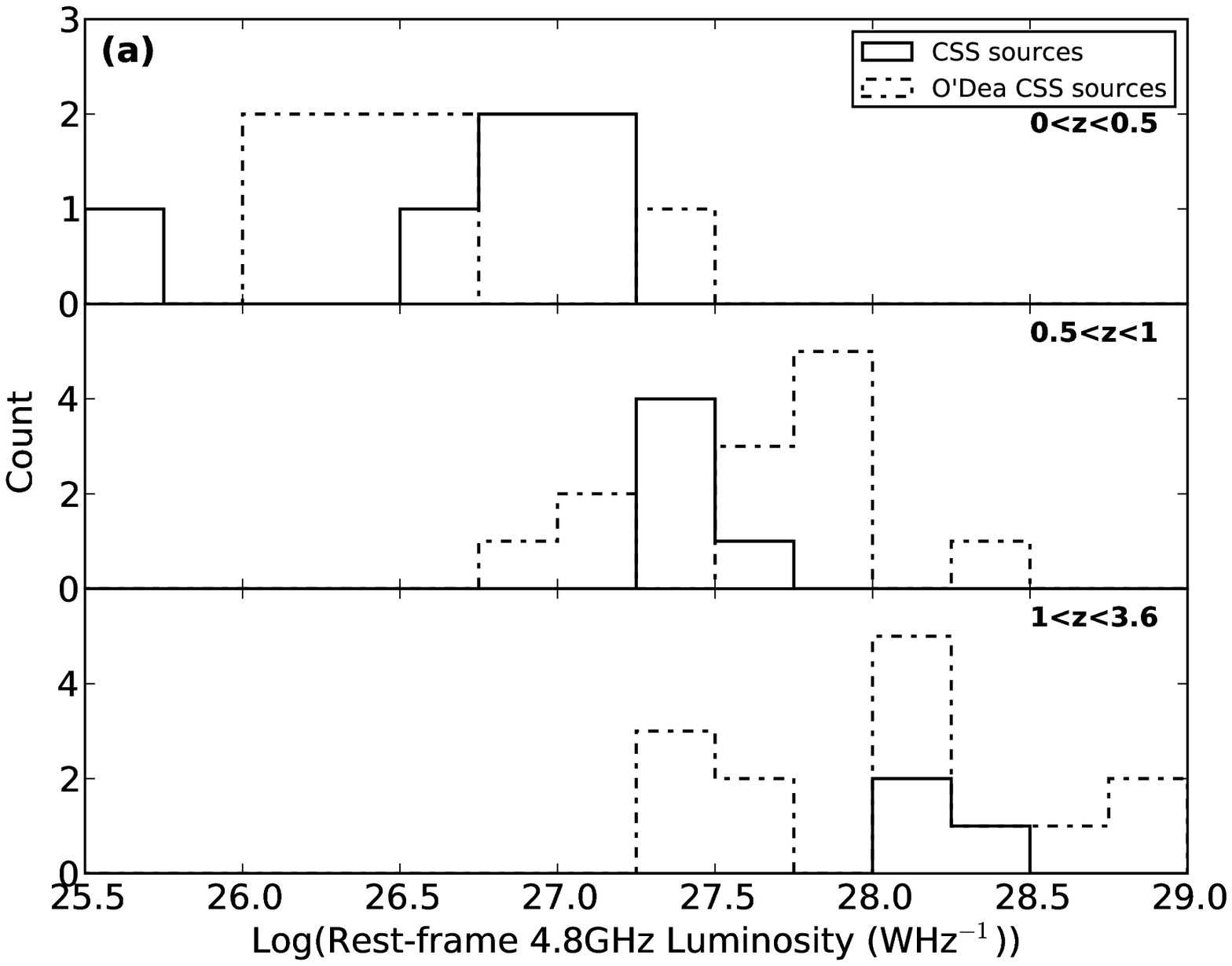}\includegraphics[scale=0.4]{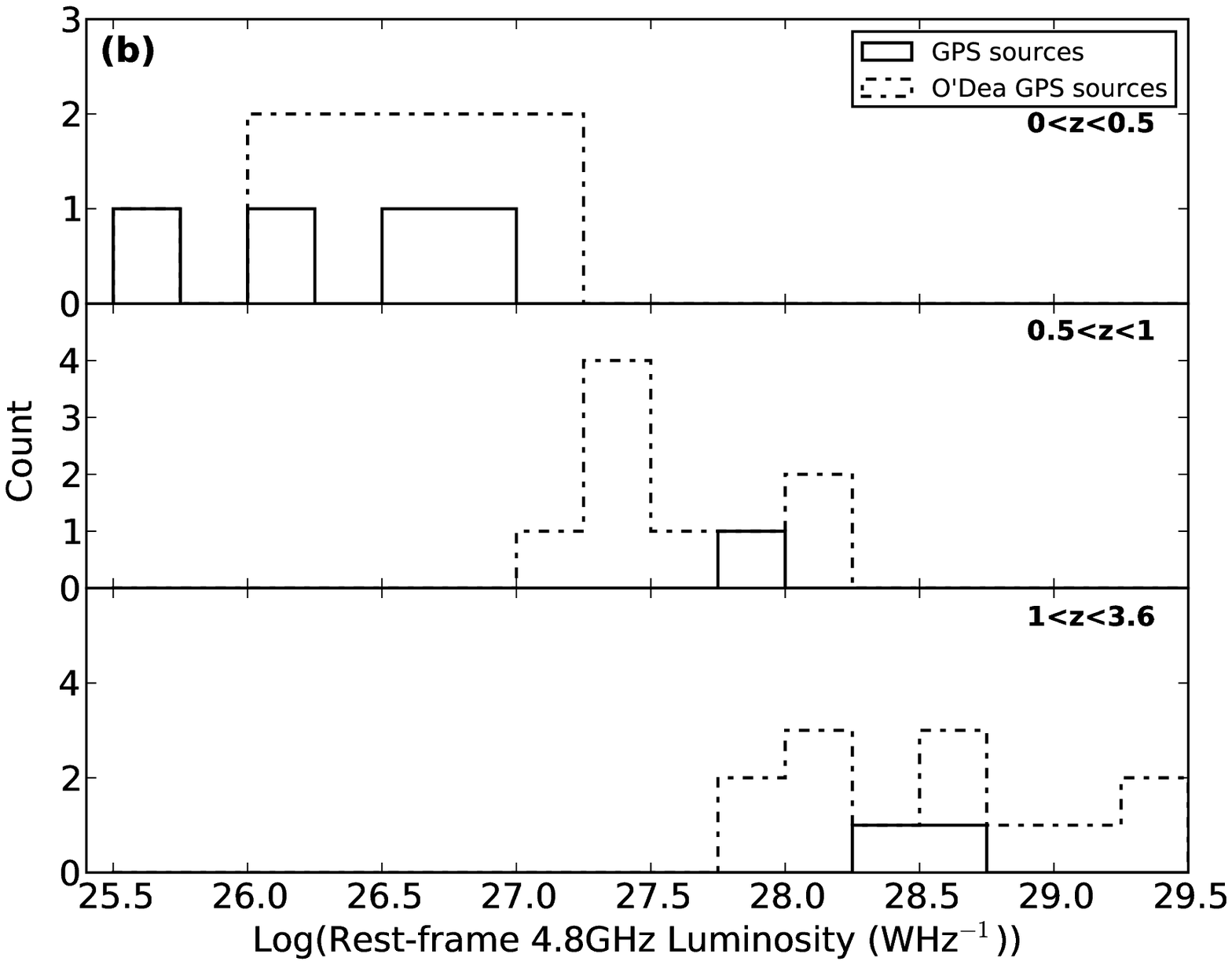}
\caption{(a) Rest-frame 5\,GHz luminosity distribution for this sample and the O'Dea sample, CSS sources, split into three redshift bins, $0<z<0.5$, $0.5<z<1$ and $1<z<3.6$, calculated using the standard $\Lambda$CDM model cosmology. (b) GPS sources for this sample and the O'Dea sample}
\label{fig:power}
\end{center}
\end{figure*}

%\textbf{Where we refer to spectral index for our sample, this indicates the spectral index $\alpha_{fit}$ across the radio spectrum covering frequencies where there are measured flux densities}. 
Distributions of redshift, spectral index and radio rest-frame luminosity are shown in Figures~\ref{fig:specind}, \ref{fig:redshift} and \ref{fig:power} respectively. In the absence of a published list of spectral indices for the O'Dea sample, we have found spectral indices by using NED to extract all the relevant radio data for each source, and measure radio spectra for all the O'Dea sample. We set a frequency range comparable to that for our sample (74\,MHz to 22\,GHz) and selected data points within this range at frequencies similar to those used for our sample. We then fitted power-law spectra using a least-squares fit, to the CSS sources, and above the peak of the GPS sources, to determine an overall spectral index ($\alpha_{fit}$) to compare to our sample. This distribution is roughly consistent with the original spectral index distribution published by O'Dea, and we use these new spectral indices for all analysis. Any discrepancies in these spectral indices are likely due to the addition of more low and high frequency radio observations, from more recent observations. We also found 6 of the 7 sources listed by O'Dea as not having a redshift to have a new spectroscopic redshift published, which we used to determine rest frame luminosities for these sources. 

\begin{table}
\begin{center}
\caption{\textsc{Summary of ATCA Observations}}
\label{table:atca}
\begin{tabular}{|cccc|} \hline
Project & Epoch & Observed &  Number of\\
& & Frequencies ($\nu_{o}$, MHz) & Sources\\
\hline
C644 & 1997 July 6--8 & 1384, 2368, 4860, 8640 & 32\\
C1813 & 2008 June 19--20 & 1384, 2368, 4860, 8640 & 52\\
\hline
\end{tabular}
\end{center}
\end{table}

The spectral index distributions for our sample and the O'Dea sample are shown in Figure~\ref{fig:specind}. The spectral indices of CSS samples between our sample and the O'Dea sample are similar, with mean spectral index $\langle\,\alpha\,\rangle=-0.77$ for our sample and $\langle\,\alpha\,\rangle=-0.70$ for the O'Dea sample. The spectral indices of our GPS sample are dominated by small number statistics, however, the mean spectral index $\langle\,\alpha\,\rangle=-0.76$ for our sample is very similar to that of the O'Dea sample ($\langle\,\alpha\,\rangle=-0.77$). Kolmogorov-Smirnov (KS) tests on the distribution of spectral indices for neither the CSS samples from the two studies, nor the GPS samples, support the hypothesis that they are drawn from different populations. The P-values are 0.32 and 0.27 respectively.

\begin{table}
%\begin{center}
\caption{\textsc{Summary of C1813 Data}}
\label{table:c1813}
\begin{tabular}{|ccccc|}\hline
Source & S$_{1.4}$ & S$_{2.3}$ & S$_{4.8}$ & S$_{8.6}$\\
 & Jy & Jy & Jy & Jy\\
\hline
PKS J0010$-$4153 & 4.43 $\pm$ 0.4 & 2.74 $\pm$ 0.3 & 1.29 $\pm$ 0.1 & 0.57 $\pm$ 0.06\\
PKS J0024$-$2928 & 2.72 $\pm$ 0.3 & 1.72 $\pm$ 0.2 & 0.99 $\pm$ 0.1 & 0.31 $\pm$ 0.03\\
PKS J0037$-$0109 & 3.84 $\pm$ 0.4 & 2.67 $\pm$ 0.3 & 1.56 $\pm$ 0.2 & 0.99 $\pm$ 0.10\\
PKS J0038$-$0207 & 5.94 $\pm$ 0.6 & 3.87 $\pm$ 0.4 & 2.21 $\pm$ 0.2 & 0.58 $\pm$ 0.06\\
PKS J0042$-$4414 & 3.62 $\pm$ 0.4 & 1.90 $\pm$ 0.2 & 0.58 $\pm$ 0.1 & 0.45 $\pm$ 0.05\\
PKS J0044$-$3530 & 2.44 $\pm$ 0.2 & 1.69 $\pm$ 0.2 & 1.01 $\pm$ 0.1 & 0.58 $\pm$ 0.06\\
PKS J0052$-$4306 & 1.96 $\pm$ 0.2 & 0.94 $\pm$ 0.1 & 0.47 $\pm$ 0.05 & 0.24 $\pm$ 0.02\\
PKS J0057$-$0123\tablenotemark{*} & 5.28 $\pm$ 0.5 & 3.83 $\pm$ 0.4 & 2.35 $\pm$ 0.2 & 1.36 $\pm$ 0.10\\
PKS J0104$-$1235 & 2.05 $\pm$ 0.2 & 1.31 $\pm$ 0.1 & 0.66 $\pm$ 0.1 & 0.31 $\pm$ 0.03\\
PKS J0108$-$1604\tablenotemark{*} & 2.77 $\pm$ 0.3 & 1.39 $\pm$ 0.1 & 0.97 $\pm$ 0.1 & 0.37 $\pm$ 0.04\\
PKS J0120$-$1520 & 4.72 $\pm$ 0.5 & 2.76 $\pm$ 0.3 & 1.29 $\pm$ 0.1 & 0.59 $\pm$ 0.06\\
PKS J0150$-$2931\tablenotemark{*} & 2.26 $\pm$ 0.2 & 1.18 $\pm$ 0.1 & 0.74 $\pm$ 0.1 & 0.46 $\pm$ 0.05\\
PKS J0200$-$3053 & 3.55 $\pm$ 0.4 & 2.03 $\pm$ 0.2 & 0.83 $\pm$ 0.1 & 0.66 $\pm$ 0.07\\
PKS J0201$-$1132 & 2.50 $\pm$ 0.3 & 1.81 $\pm$ 0.2 & 1.36 $\pm$ 0.1 & 1.07 $\pm$ 0.10\\
PKS J0215$-$1259 & 3.29 $\pm$ 0.3 & 1.81 $\pm$ 0.2 & 1.34 $\pm$ 0.1 & 0.48 $\pm$ 0.05\\
PKS J0220$-$0156 & 3.24 $\pm$ 0.3 & 1.91 $\pm$ 0.2 & 0.88 $\pm$ 0.1 & 0.40 $\pm$ 0.04\\
PKS J0237$-$1932\tablenotemark{*} & 3.67 $\pm$ 0.4 & 1.89 $\pm$ 0.2 & 0.92 $\pm$ 0.1 & 0.58 $\pm$ 0.06\\
PKS J0321$-$4510\tablenotemark{*} & 3.03 $\pm$ 0.3 & 1.90 $\pm$ 0.2 & 0.74 $\pm$ 0.1 & 0.16 $\pm$ 0.02\\
PKS J0407$-$1211\tablenotemark{*} & 2.75 $\pm$ 0.3 & 1.89 $\pm$ 0.2 & 1.21 $\pm$ 0.1 & 1.16 $\pm$ 0.10\\
PKS J0408$-$7507 & 13.13 $\pm$ 1.0 & 7.71 $\pm$ 0.8 & 2.40 $\pm$ 0.2 & 1.73 $\pm$ 0.20\\
PKS J0444$-$2809\tablenotemark{*} & 6.78 $\pm$ 0.7 & 4.31 $\pm$ 0.4 & 1.27 $\pm$ 0.1 & 0.64 $\pm$ 0.06\\
PKS J0455$-$3006\tablenotemark{*} & 3.31 $\pm$ 0.3 & 2.17 $\pm$ 0.2 & 1.21 $\pm$ 0.1 & 0.69 $\pm$ 0.07\\
PKS J0458$-$3007\tablenotemark{*} & 2.58 $\pm$ 0.3 & 1.79 $\pm$ 0.2 & 0.85 $\pm$ 0.1 & 0.28 $\pm$ 0.03\\
PKS J0621$-$5241\tablenotemark{*} & 3.17 $\pm$ 0.3 & 2.20 $\pm$ 0.2 & 0.35 $\pm$ 0.04 & 0.26 $\pm$ 0.03\\
PKS J0626$-$5341\tablenotemark{*} & 6.62 $\pm$ 0.7 & 3.89 $\pm$ 0.4 & 1.84 $\pm$ 0.2 & 0.90 $\pm$ 0.09\\
PKS J1555$-$7940\tablenotemark{*} & 3.51 $\pm$ 0.4 & 1.54 $\pm$ 0.2 & 0.72 $\pm$ 0.1 & 0.47 $\pm$ 0.05\\
PKS J1724$-$0242\tablenotemark{*} & 2.19 $\pm$ 0.2 & 1.81 $\pm$ 0.2 & 4.56 $\pm$ 0.5 & 0.27 $\pm$ 0.03\\
PKS J1937$-$5838 & 2.87 $\pm$ 0.3 & 1.81 $\pm$ 0.2 & 0.99 $\pm$ 0.1 & 0.54 $\pm$ 0.05\\
PKS J1941$-$1524 & 6.63 $\pm$ 0.7 & 4.39 $\pm$ 0.4 & 1.83 $\pm$ 0.2 & 0.67 $\pm$ 0.07\\
PKS J2006$-$0223 & 2.10 $\pm$ 0.2 & 1.58 $\pm$ 0.2 & 1.04 $\pm$ 0.1 & 0.60 $\pm$ 0.06\\
PKS J2024$-$5723\tablenotemark{*} & 2.04 $\pm$ 0.2 & 1.11 $\pm$ 0.1 & 0.61 $\pm$ 0.1 & 0.28 $\pm$ 0.03\\
PKS J2035$-$3454\tablenotemark{*} & 5.55 $\pm$ 0.6 & 3.35 $\pm$ 0.3 & 1.79 $\pm$ 0.2 & 0.87 $\pm$ 0.09\\
PKS J2056$-$1956\tablenotemark{*} & 2.64 $\pm$ 0.3 & 1.47 $\pm$ 0.1 & 0.98 $\pm$ 0.1 & 0.52 $\pm$ 0.05\\
PKS J2137$-$1433\tablenotemark{*} & 2.40 $\pm$ 0.2 & 2.17 $\pm$ 0.2 & 0.53 $\pm$ 0.1 & 0.12 $\pm$ 0.01\\
PKS J2143$-$4312 & 2.66 $\pm$ 0.3 & 1.80 $\pm$ 0.2 & 0.52 $\pm$ 0.1 & 0.30 $\pm$ 0.03\\
PKS J2147$-$8132\tablenotemark{*} & 2.86 $\pm$ 0.3 & 1.65 $\pm$ 0.2 & 0.57 $\pm$ 0.1 & 0.31 $\pm$ 0.03\\
PKS J2214$-$1701\tablenotemark{*} & 9.05 $\pm$ 0.9 & 5.21 $\pm$ 0.5 & 2.37 $\pm$ 0.2 & 1.07 $\pm$ 0.10\\
PKS J2253$-$4057 & 3.05 $\pm$ 0.3 & 1.41 $\pm$ 0.1 & 0.84 $\pm$ 0.1 & 0.40 $\pm$ 0.04\\
PKS J2255$-$5245 & 2.69 $\pm$ 0.3 & 1.79 $\pm$ 0.2 & 0.92 $\pm$ 0.1 & 0.50 $\pm$ 0.05\\
PKS J2319$-$2728\tablenotemark{*} & 2.77 $\pm$ 0.3 & 1.89 $\pm$ 0.2 & 1.13 $\pm$ 0.1 & 0.66 $\pm$ 0.07\\
PKS J2326$-$0202\tablenotemark{*} & 2.34 $\pm$ 0.2 & 1.67 $\pm$ 0.2 & 0.24 $\pm$ 0.02 & 0.23 $\pm$ 0.02\\
PKS J2326$-$4027 & 3.26 $\pm$ 0.3 & 2.06 $\pm$ 0.2 & 1.07 $\pm$ 0.1 & 0.54 $\pm$ 0.05\\
\hline
\end{tabular}\\
\textsc{Notes.}-- $^*$ indicates this source was not part among the 76 candidate CSS sources. All errors are due to inherent systematic errors in the flux calibration and are $\sim10\%$.
%\end{center}
\end{table}

\begin{landscape}
\begin{center}
\begin{deluxetable}{|lllllllllllllllll|}
\tablewidth{0pt}
\tabletypesize{\scriptsize}
\tablecaption{\textsc{Table of Data for CSS and GPS candidates and sources}
\label{table:sample}}
\tablehead{
\colhead{\footnotesize Source \& Tag\tablenotemark{1}} &\colhead{\footnotesize RA} & \colhead{\footnotesize Dec.} & \colhead{\footnotesize R mag.} & \colhead{\footnotesize Redshift} & \colhead{\footnotesize Type} & \colhead{\footnotesize $\alpha_{fit}$\tablenotemark{2}} & \colhead{\footnotesize $\Delta\,\alpha_{fit}$} & \colhead{\footnotesize $\alpha^{5}_{2.7}$}  &
\colhead{\footnotesize Physical} & \colhead{\footnotesize} & \colhead{\footnotesize} & \colhead{\footnotesize PMN} & \colhead{\footnotesize} & \colhead{\footnotesize AT20G} & \colhead{\footnotesize References}\\
 & & & & & & & & & \colhead{\footnotesize Size} & \colhead{\footnotesize 1.4\,GHz} & \colhead{\footnotesize 2.7\,GHz} &  \colhead{\footnotesize 5\,GHz} & \colhead{\footnotesize 8.4\,GHz} & \colhead{\footnotesize 20\,GHz} & \\
 & & & & & & & & & \colhead{\footnotesize kpc} & \colhead{\footnotesize Jy} &  \colhead{\footnotesize Jy} & \colhead{\footnotesize Jy} & \colhead{\footnotesize Jy} & \colhead{\footnotesize Jy} &\\
}
\startdata
\footnotesize{PKS J0024$-$4202 G,k}&\footnotesize{00 24 42.95}&\footnotesize{$-$42 02 03.5}&\footnotesize{20.2}&\footnotesize{0.937}&\footnotesize{G}&\footnotesize{$-$0.92}&\footnotesize{0.008}&\footnotesize{$-$0.77}&\footnotesize{15.0}&\footnotesize{2.78\tablenotemark{*}}&\footnotesize{2.93\tablenotemark{*}}&\footnotesize{1.85\tablenotemark{*}}&\footnotesize{0.91\tablenotemark{*}}&\footnotesize{0.29}&\footnotesize{1,2,3}\\
\footnotesize{PKS J0025$-$2602 C,k}&\footnotesize{00 25 49.18}&\footnotesize{$-$26 02 12.7}&\footnotesize{18.8}&\footnotesize{0.322}&\footnotesize{G}&\footnotesize{$-$0.54}&\footnotesize{0.02}&\footnotesize{$-$0.70}&\footnotesize{3.2}&\footnotesize{8.37\tablenotemark{*}}&\footnotesize{5.94\tablenotemark{*}}&\footnotesize{3.64\tablenotemark{*}}&\footnotesize{2.15\tablenotemark{*}}&\footnotesize{0.98}&\footnotesize{4}\\
\footnotesize{PKS J0116$-$2052 C,k}&\footnotesize{01 16 51.47}&\footnotesize{$-$20 52 06.5}&\footnotesize{21.6}&\footnotesize{1.41}&\footnotesize{G}&\footnotesize{$-$0.76}&\footnotesize{0.02}&\footnotesize{$-$1.07}&\footnotesize{6.0}&\footnotesize{3.81\tablenotemark{*}}&\footnotesize{2.44\tablenotemark{*}}&\footnotesize{1.27\tablenotemark{*}}&\footnotesize{0.64\tablenotemark{*}}&\footnotesize{0.24}&\footnotesize{2,5,6}\\
\footnotesize{PKS J0203$-$4349 C,k}&\footnotesize{02 03 40.75}&\footnotesize{$-$43 49 50.9}&\footnotesize{18.7}&\footnotesize{0.45\tablenotemark{p}}&\footnotesize{G}&\footnotesize{$-$0.73}&\footnotesize{0.01}&\footnotesize{$-$0.85}&\footnotesize{8.6}&\footnotesize{2.67\tablenotemark{*}}&\footnotesize{1.81\tablenotemark{*}}&\footnotesize{1.03\tablenotemark{*}}&\footnotesize{0.58\tablenotemark{*}}&\footnotesize{0.23}&\footnotesize{1}\\
\footnotesize{PKS J0240$-$2309 G,k}&\footnotesize{02 40 08.13}&\footnotesize{$-$23 09 15.7}&\footnotesize{16.1}&\footnotesize{2.223}&\footnotesize{Q}&\footnotesize{$-$0.52}&\footnotesize{0.009}&\footnotesize{$-$0.76}&\footnotesize{8.5}&\footnotesize{4.4}&\footnotesize{2.41}&\footnotesize{3.34}&\footnotesize{2.2}&\footnotesize{0.90}&\footnotesize{3,7,8}\\
\footnotesize{PKS J0252$-$7104 C,k}&\footnotesize{02 52 46.29}&\footnotesize{$-$71 04 33.9}&\footnotesize{19.2}&\footnotesize{0.568}&\footnotesize{G}&\footnotesize{$-$1.02}&\footnotesize{0.006}&\footnotesize{$-$1.14}&\footnotesize{1.6}&\footnotesize{5.59\tablenotemark{*}}&\footnotesize{3.42\tablenotemark{*}}&\footnotesize{1.69\tablenotemark{*}}&\footnotesize{0.80\tablenotemark{*}}&\footnotesize{0.32}&\footnotesize{1,4}\\
\footnotesize{PKS J0420$-$6223 C,k}&\footnotesize{04 20 56.17}&\footnotesize{$-$62 23 38.2}&\footnotesize{$>$21.5}&\footnotesize{0.81\tablenotemark{p}}&\footnotesize{G}&\footnotesize{$-$1.10}&\footnotesize{0.005}&\footnotesize{$-$1.12}&\footnotesize{15.9}&\footnotesize{3.26\tablenotemark{*}}&\footnotesize{1.94\tablenotemark{*}}&\footnotesize{0.93\tablenotemark{*}}&\footnotesize{0.47\tablenotemark{*}}&\footnotesize{0.17}&\footnotesize{5}\\
\footnotesize{PKS J0616$-$3456 C,k}&\footnotesize{06 16 35.97}&\footnotesize{$-$34 56 17.6}&\footnotesize{18.7}&\footnotesize{0.329}&\footnotesize{G}&\footnotesize{$-$0.56}&\footnotesize{0.006}&\footnotesize{$-$0.63}&\footnotesize{8.9}&\footnotesize{2.7}&\footnotesize{1.96}&\footnotesize{1.31}&\footnotesize{0.78}&\footnotesize{0.44}&\footnotesize{1,6,7}\\
\footnotesize{PKS J0943$-$0819 G,k}&\footnotesize{09 43 37.01}&\footnotesize{$-$08 19 30.7}&\footnotesize{17.9}&\footnotesize{0.228}&\footnotesize{G}&\footnotesize{$-$0.78}&\footnotesize{0.001}&\footnotesize{$-$0.79}&\footnotesize{3.8}&\footnotesize{2.48}&\footnotesize{1.73}&\footnotesize{1.12}&\footnotesize{0.75}&\footnotesize{0.33}&\footnotesize{9,10}\\
\footnotesize{PKS J1154$-$3505 G,k}&\footnotesize{11 54 21.79}&\footnotesize{$-$35 05 29.1}&\footnotesize{17.8}&\footnotesize{0.258}&\footnotesize{G}&\footnotesize{$-$0.69}&\footnotesize{0.001}&\footnotesize{$-$0.66}&\footnotesize{0.7}&\footnotesize{6.1}&\footnotesize{4.12}&\footnotesize{2.56}&\footnotesize{1.92}&\footnotesize{0.89}&\footnotesize{1}\\
\footnotesize{PKS J1218$-$4600 C,k}&\footnotesize{12 18 06.26}&\footnotesize{$-$46 00 30.2}&\footnotesize{18.9}&\footnotesize{0.529}&\footnotesize{Q}&\footnotesize{$-$0.53}&\footnotesize{0.007}&\footnotesize{$-$0.82}&\footnotesize{6.9}&\footnotesize{4.6}&\footnotesize{3.3}&\footnotesize{2.29}&\footnotesize{1.27}&\footnotesize{0.70}&\footnotesize{1,2,7}\\
\footnotesize{PKS J1223$-$4235 C,k}&\footnotesize{12 23 43.40}&\footnotesize{ $-$42 35 32.2}&\footnotesize{15.7}&\footnotesize{0.176}&\footnotesize{G}&\footnotesize{$-$0.63}&\footnotesize{0.03}&\footnotesize{$-$0.78}&\footnotesize{6.2}&\footnotesize{2.73}&\footnotesize{1.62}&\footnotesize{1.10}&\footnotesize{0.63}&\footnotesize{0.33}&\footnotesize{1,11,12,13}\\
\footnotesize{PKS J1248$-$1959 G,k}&\footnotesize{12 48 23.87}&\footnotesize{$-$19 59 18.4}&\footnotesize{20.3}&\footnotesize{1.275}&\footnotesize{Q}&\footnotesize{$-$0.84}&\footnotesize{0.0005}&\footnotesize{$-$0.76}&\footnotesize{12.6}&\footnotesize{4.9}&\footnotesize{3.94}&\footnotesize{2.42}&\footnotesize{1.36}&\footnotesize{0.69}&\footnotesize{6,14}\\
\footnotesize{PKS J1308$-$0950 C,k}&\footnotesize{13 08 39.06}&\footnotesize{$-$09 50 32.1}&\footnotesize{18.7}&\footnotesize{0.464}&\footnotesize{G}&\footnotesize{$-$0.54}&\footnotesize{0.004}&\footnotesize{$-$0.69}&\footnotesize{2.7}&\footnotesize{4.26}&\footnotesize{2.88}&\footnotesize{1.88}&\footnotesize{1.11}&\footnotesize{0.73}&\footnotesize{1}\\
\footnotesize{PKS J1311$-$2216 C}&\footnotesize{13 11 39.37}&\footnotesize{$-$22 16 41.4}&\footnotesize{...}&\footnotesize{0.8}&\footnotesize{Q}&\footnotesize{$-$1.04}&\footnotesize{0.01}&\footnotesize{$-$1.30}&\footnotesize{15.8}&\footnotesize{5.24}&\footnotesize{2.43}&\footnotesize{1.24}&\footnotesize{0.45}&\footnotesize{0.19}&\footnotesize{5,14}\\
\footnotesize{PKS J1323$-$4452 C,k}&\footnotesize{13 23 04.25}&\footnotesize{$-$44 52 33.1}&\footnotesize{18.6}&\footnotesize{1.95}&\footnotesize{Q}&\footnotesize{$-$0.86}&\footnotesize{0.007}&\footnotesize{$-$0.88}&\footnotesize{17.0}&\footnotesize{3.47}&\footnotesize{1.79}&\footnotesize{1.06}&\footnotesize{0.52}&\footnotesize{0.25}&\footnotesize{1}\\
\footnotesize{PKS J1526$-$1351 C,k}&\footnotesize{15 26 59.47}&\footnotesize{$-$13 51 00.1}&\footnotesize{19.6}&\footnotesize{1.687}&\footnotesize{Q}&\footnotesize{$-$0.52}&\footnotesize{0.009}&\footnotesize{$-$0.58}&\footnotesize{3.4}&\footnotesize{2.4}&\footnotesize{1.7}&\footnotesize{1.31}&\footnotesize{0.82}&\footnotesize{0.39}&\footnotesize{7,15}\\
\footnotesize{PKS J1744$-$5144 G,k}&\footnotesize{17 44 25.43}&\footnotesize{$-$51 44 45.7}&\footnotesize{20.8}&\footnotesize{0.63\tablenotemark{p}}&\footnotesize{G}&\footnotesize{$-$0.57}&\footnotesize{0.003}&\footnotesize{$-$0.72}&\footnotesize{0.4}&\footnotesize{...}&\footnotesize{4.6}&\footnotesize{3.90}&\footnotesize{2.37}&\footnotesize{1.24}&\footnotesize{1,2,16,27}\\
\footnotesize{PKS J1818$-$5158 C,k}&\footnotesize{18 18 07.11}&\footnotesize{$-$51 58 11.1}&\footnotesize{...}&\footnotesize{0.48\tablenotemark{p}}&\footnotesize{Q}&\footnotesize{$-$1.30}&\footnotesize{0.005}&\footnotesize{$-$1.51}&\footnotesize{16.1}&\footnotesize{3.5}&\footnotesize{1.6}&\footnotesize{0.72}&\footnotesize{0.38}&\footnotesize{0.07}&\footnotesize{1}\\
\footnotesize{PKS J1819$-$6345 C,k}&\footnotesize{18 19 34.99}&\footnotesize{$-$63 45 48.2}&\footnotesize{15.6}&\footnotesize{0.0627}&\footnotesize{G}&\footnotesize{$-$0.86}&\footnotesize{0.005}&\footnotesize{$-$0.92}&\footnotesize{0.5}&\footnotesize{13.34\tablenotemark{*}}&\footnotesize{8.80\tablenotemark{*}}&\footnotesize{5.14\tablenotemark{*}}&\footnotesize{2.95\tablenotemark{*}}&\footnotesize{1.71}&\footnotesize{1,8,18}\\
\footnotesize{PKS J1830$-$3602 G,k}&\footnotesize{18 30 58.88}&\footnotesize{$-$36 02 30.3}&\footnotesize{14.1}&\footnotesize{0.12\tablenotemark{p}}&\footnotesize{G}&\footnotesize{$-$1.47}&\footnotesize{0.003}&\footnotesize{$-$1.34}&\footnotesize{3.4}&\footnotesize{7.12\tablenotemark{*}}&\footnotesize{3.54\tablenotemark{*}}&\footnotesize{1.32\tablenotemark{*}}&\footnotesize{0.51\tablenotemark{*}}&\footnotesize{0.14}&\footnotesize{1}\\
\footnotesize{PKS J1939$-$6342 G,k}&\footnotesize{19 39 24.83}&\footnotesize{$-$63 42 45.3}&\footnotesize{17.5}&\footnotesize{0.183}&\footnotesize{G}&\footnotesize{$-$1.30}&\footnotesize{$2\times\,10^{-6}$}&\footnotesize{$-$1.02}&\footnotesize{0.2}&\footnotesize{16.4}&\footnotesize{11.5}&\footnotesize{5.66}&\footnotesize{3}&\footnotesize{0.96}&\footnotesize{4,7}\\
\footnotesize{PKS J1957$-$4222 C,k}&\footnotesize{19 57 15.31}&\footnotesize{$-$42 22 20.1}&\footnotesize{16.5}&\footnotesize{0.82\tablenotemark{p}}&\footnotesize{Q}&\footnotesize{$-$0.93}&\footnotesize{0.009}&\footnotesize{$-$0.95}&\footnotesize{15.2}&\footnotesize{3.25\tablenotemark{*}}&\footnotesize{1.90\tablenotemark{*}}&\footnotesize{0.92\tablenotemark{*}}&\footnotesize{0.43\tablenotemark{*}}&\footnotesize{0.15}&\footnotesize{1}\\
\footnotesize{PKS J2011$-$0644 G,k}&\footnotesize{20 11 14.23}&\footnotesize{$-$06 44 03.4}&\footnotesize{21.5}&\footnotesize{0.547}&\footnotesize{G}&\footnotesize{$-$0.74}&\footnotesize{0.002}&\footnotesize{$-$0.73}&\footnotesize{0.2}&\footnotesize{2.60}&\footnotesize{2.09}&\footnotesize{1.44}&\footnotesize{0.77}&\footnotesize{0.37}&\footnotesize{9,19}\\
\footnotesize{PKS J2137$-$2042 C,k}&\footnotesize{21 37 50.02}&\footnotesize{ $-$20 42 31.5}&\footnotesize{19.3}&\footnotesize{0.636}&\footnotesize{G}&\footnotesize{$-$0.57}&\footnotesize{0.02}&\footnotesize{$-$0.82}&\footnotesize{1.7}&\footnotesize{...}&\footnotesize{2.49}&\footnotesize{1.55}&\footnotesize{0.76}&\footnotesize{0.41}&\footnotesize{4,8}\\
\footnotesize{PKS J2152$-$2828 C}&\footnotesize{21 52 03.79}&\footnotesize{$-$28 28 28.7}&\footnotesize{19.2}&\footnotesize{0.479}&\footnotesize{G}&\footnotesize{$-$0.59}&\footnotesize{0.007}&\footnotesize{$-$0.67}&\footnotesize{11.9}&\footnotesize{2.75\tablenotemark{*}}&\footnotesize{1.98\tablenotemark{*}}&\footnotesize{1.25\tablenotemark{*}}&\footnotesize{0.75\tablenotemark{*}}&\footnotesize{0.37}&\footnotesize{5}\\
\enddata\\
\textsc{Notes.}-- The full version of this table is available in the online version of this paper as Supporting Information. All flux densities listed are from observer frame frequencies and this is listed for each catalogue. $^1$Tag for CSS or GPS candidate or source in this catalogue, k indicates already catalogued as a CSS or GPS source, $^P$Photometric redshift from Burgess \& Hunstead 2006 (all other redshifts are spectroscopic), $^2$$\alpha_{fit}$ is the spectral index given by our least squares spectral fit across the existing data for each source. For GPS sources, this spectral index is the spectral index above the peak of the radio spectrum. The error on $\alpha_{fit}$, given by the sum of the residuals of the least-squares power-law spectral fit divided by the number of data points is given in the next column. $^*$ Indicates that this flux density is new and previously unpublished, from C1813 (See \S~\ref{sec:c1813}); otherwise the flux density measurements are from PKSCAT90. In the case of C1813 measurements in Column 12 (2.7\,GHz), the flux densities were measured at 2.3\,GHz, whereas PKSCAT90 flux densities were measured at 2.7\,GHz. All other flux densities are from the catalogues indicated. Column (5) refers to Galaxy (G) or quasar (Q) or empty field (EF) in the SuperCOSMOS science archive.\\
\textsc{References.}-- (1) Burgess \& Hunstead 2006, (2) di Serego-Alighieri et al. 1994, (3) de Vries, Barthel \& Hes 1995, (4) Tadhunter et al. 1993, (5) McCarthy et al. 1996, (6) Kapahi et al. 1998, (7) Edwards \& Tingay 2004, (8) Holt et al. 2008, (9) O'Dea 1998, (10) O'Dea \& Baum 1997, (11) Simpson et al. 1993, (12) Safouris, Hunstead \& Prouton 2003, (13) Johnston et al. 2005, (14) Costa 2002, (15) Hewitt \& Burbidge 1989, (16) Labiano et al. 2007, (17) Jauncey et al. 2003, (18) Danziger \& Goss 1979, (19) Snellen et al. 2002.
\end{deluxetable}
\end{center}
\end{landscape}

The redshift distributions for the current sample of CSS and GPS sources are dominated by sources at lower redshift compared to the O'Dea sample (Figure~\ref{fig:redshift}). The redshifts of CSS sources between our sample and the O'Dea sample are dissimilar, with a mean redshift $\langle\,z\rangle=0.70$ for our sample and $\langle\,z\rangle=1.03$ for the O'Dea sample. A KS test marginally supports the hypothesis that they are drawn from different populations (P=0.04).

Although our GPS sample suffers from small number statistics, the mean redshift $\langle\,z\rangle=0.71$ for our sample is again dissimilar to that of the O'Dea sample ($\langle\,z\rangle=1.06$). A KS test does not support the hypothesis that they are drawn from different populations (P=0.50). Our sample appears more homogenous in redshift spread than the O'Dea sample, although our sources are on average at lower redshifts.
\begin{table*}
\begin{center}
\caption{\textsc{Comparison of statistics from the O'Dea sample and our sample}}
\label{table:means}
\begin{tabular}{|ccccccccc|} \hline
Source Type & O'Dea Sample & This Sample & O'Dea Sample & This Sample & O'Dea Sample & This Sample & O'Dea Sample & This Sample \\
& Mean & Mean & Median & Median & Mean & Mean & Median & Median\\
& Spectral Index & Spectral Index & Spectral Index & Spectral Index & Redshift & Redshift & Redshift & Redshift\\
\hline
All sources & $-0.73$ & $-0.76$ & $-0.74$ & $-0.69$ & 1.05 & 0.71 & 0.86 & 0.54\\
CSS & $-0.70$ & $-0.77$ & $-0.66$ & $-0.73$ & 1.03 & 0.70 & 0.88 & 0.53\\
GPS & $-0.77$ & $-0.76$ & $-0.78$ & $-0.65$ & 1.06 & 0.71 & 0.74 & 0.55\\
Galaxies & $-0.73$ & $-0.75$ & $-0.74$ & $-0.69$ & 0.71 & 0.48 & 0.55 & 0.46\\
QSOs & $-0.73$ & $-0.80$ & $-0.73$ & $-0.74$ & 1.38 & 1.22 & 1.12 & 1.05\\
CSS Galaxies & $-0.59$ & $-0.71$ & $-0.57$ & $-0.63$ & 0.77 & 0.52 & 0.69 & 0.46\\
GPS Galaxies & $-0.83$ & $-0.81$ & $-0.81$ & $-0.74$ & 0.66 & 0.41 & 0.47 & 0.26\\
CSS Quasars & $-0.76$ & $-0.86$ & $-0.73$ & $-0.90$ & 1.22 & 1.04 & 1.06 & 0.81\\
GPS Quasars & $-0.68$ & $-0.60$ & $-0.67$ & $-0.60$ & 1.61 & 1.75 & 1.30 & 1.75\\
\hline
\end{tabular}
\textsc{Notes.}-- Standard deviations for the samples are $\sim0.3$ for all spectral index measurements, and $\sim0.5$ for all redshift measurements.
\end{center}
\end{table*}
To determine whether our lower mean redshifts arise from selection effects, we compare the radio luminosities of both samples (Figure~\ref{fig:power}). We also show luminosity against redshift (Figure~\ref{fig:lumz}). The distributions of radio luminosity are split by redshift into three bins: $0<z<0.5$, $0.5<z<1$ and $1<z<3.6$. The O'Dea sources are more luminous than our sample in all redshift ranges, but not by any significant amount. This would have been due to the lack of a minimum power cut in our source selection, and hence we may include lower luminosity sources excluded by the power threshold in the O'Dea sample. Our distribution peaks at $L_{5}\sim10^{27}$\,WHz$^{-1}$, for the CSS sources, whereas the O'Dea sample peaks at around $L_{5}=10^{28}$\,WHz$^{-1}$. Due to the small number of GPS sources in our sample, we cannot reliably determine a peak in their luminosity distribution. The O'Dea GPS candidates have a small peak at $L_{5}\sim10^{28}$\,WHz$^{-1}$, but the overall distribution is quite flat. Unlike the O'Dea sample, our GPS candidates do not appear to be more luminous than the CSS candidates. No strong evidence exists that Doppler boosting has a significant effect on the radio emission for these objects \citep{Review}, and we assume the radio power here to be intrinsic. 

KS tests on the CSS and GPS luminosities give P=0.02 and P=0.78 respectively. Although the results for the CSS source redshift and luminosity distributions offer marginal support for our CSS sample being drawn from a different population to the O'Dea sample, we note that neither sample is complete. There has been no previous indication or expectation that samples drawn from the northern or southern hemispheres would be different, and this marginal result is likely due to the incompleteness of both samples. We have also performed KS tests on the 5\,GHz flux density and R-band magnitudes, for CSS and GPS sources, giving P=0.45 and P=0.97 for the flux densities respectively and P=0.35 and P=0.69 for the R-band magnitude respectively. Together these comparisons suggest that the bulk of the physical properties are consistent with our samples having been drawn from the same underlying population as those presented by O'Dea. This result should be explored in more detail, however, with a complete sample selected in a homogenous manner, such as the algorithm we are suggesting in this paper.

We also note that the numbers of sources selected from our parent sample are quite small (0.3\% of the 8264 PKSCAT90 sources, 12\% of the 218 sources before the steep and compact selection criteria were applied, and 34\% of the initial sample of 76 sources), compared to the O'Dea sample (20\% and 6\% of CSS and GPS sources respectively from the relevant parent samples). Our selection rejects certain sources already classified as CSS or GPS, and we have discussed possible reasons for the exclusion of the four of the O'Dea GPS sources in our sample. It may be primarily the selection frequency which removes such sources, or the strict size criteria.

\subsection{Luminosity Effects}
\label{sec:evolution}
\begin{figure}
\begin{center}
\includegraphics[scale=0.4]{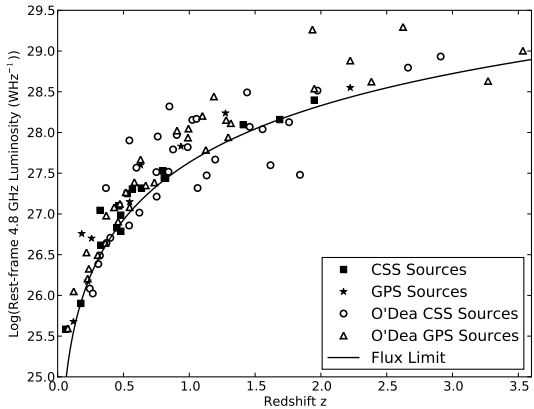}
\caption{Rest-frame 5\,GHz luminosities for our entire sample and the sample from O'Dea, versus redshift. The lowest redshift CSS source from our sample is the blue disk galaxy discussed in \S~\ref{sec:coolsources}.}
\label{fig:lumz}
\end{center}
\end{figure}

We have investigated both our sample and the O'Dea sample to determine whether we see any evolution with redshift.  We can see in Figure~\ref{fig:lumz} that there are no extremely luminous objects at low redshift ($z<0.3$), to which our selection would have been sensitive if they exist. We use the $1/V_{max}$ method to calculate space densities of these objects at higher redshift and to determine whether we could be missing luminous nearby sources. Figure~\ref{fig:lumz} gives the rest-frame luminosities at 5\,GHz (using our measured spectral indices for the O'Dea sample) in standard $\Lambda$CDM cosmology.  

We choose several different redshift bins, at medium to high redshift, to perform the density calculations.  The  $1/V_{max}$ method is used to find the space density $\Phi = \sum \limits_{i=1}^N 1/V_{max}$.  We calculated $1/V_{max}$ for several different redshift bins, taking into account our flux limit of S$_{\nu_{o}=2.7}> 1.5$\,Jy and the volume differences between the samples.
We then compare these values to the volume of the universe out to a redshift $z=0.3$, of $V=1.53\times\,10^{10}\,Mpc^3$, to estimate the number of expected nearby luminous sources, in the absence of evolutionary effects. From the calculations, using both our sample, and the O'Dea sample separately and combined (see Table~\ref{table:spdens}), we confirm that we expect to see $<1$ CSS or GPS source with luminosity $L_{\nu_{r}=5}>10^{27}$WHz$^{-1}$ in the volume out to $z=0.3$. This confirms we are not missing any nearby extremely luminous objects from our sample. 

Another effect is that our sample appears to select less luminous sources, and at lower redshift, compared to the O'Dea sample (\S~\ref{sec:comparison}. Our luminosity distributions have several minor differences (Figure~\ref{fig:power}), and we do not see the very luminous, high redshift sources in our sample that are contained within O'Dea's sample, although they do not seem to be drawn from different populations. The O'Dea selection criteria are heterogeneous, and objects are drawn from catalogues with different frequencies. 

It may be that we do not select the very luminous GPS sources such as those seen in the O'Dea sample in our sample for several reasons. If we have a source selected with a steep spectrum at low frequencies, (such as 178\,MHz, the frequency used for selecting the CSS sources in the O'Dea sample), the spectral index may flatten to a value of $\alpha>-0.5$ at higher frequencies (such as 5\,GHz), thus failing to meet our selection criteria. If the source spectrum does flatten at higher frequencies, it will result in a more luminous source at that redshift, due to the higher flux density. Similarly, if the radio spectra flatten out for these objects at lower frequencies (if the spectrum is starting to show a spectral turnover), they may not be selected using our criteria. This suggests it is possible that a high frequency flux limit (as for our sample) will remove any low frequency luminous sources, as in the O'Dea GPS sample. The four luminous OÕDea GPS sources not selected in our sample do not have a spectral index $\alpha^{2.7}_{5}>-0.5$, which suggests that GPS sources in particular, are high redshift, highly luminous objects, with flatter radio spectra at high frequencies. From our comparison of these two samples, we note that particular choices of selection criteria may end up unintentionally biasing against sources of interest. It is clear that the selection criteria for CSS and GPS sources must remain general enough not to exclude such sources, while still minimising the numbers of non-CSS/GPS sources selected. These criteria would need to be applied to large homogenous samples to select all actual CSS/GPS candidate sources, in order to to find the fraction of real GPS and CSS sources.
	
\begin{table}
\caption{\textsc{Summary of space densities for various redshifts bins}}
\label{table:spdens}
\begin{tabular}{|ccccc|} \hline
Redshift & $\Phi$ & $\Phi_{O'Dea}$ & $\Phi_{combined}$ & \% Error\\
\hline
0.6 - 1.0 & 3.36$\times\,10^{-11}$ & 4.66$\times\,10^{-11}$ & 8.02$\times\,10^{-11}$ & 22\%\\
0.6 - 1.3 & 3.40$\times\,10^{-11}$ & 6.05$\times\,10^{-11}$ & 9.45$\times\,10^{-11}$ & 18\%\\
0.6 - 1.5 & 3.47$\times\,10^{-11}$ & 6.14$\times\,10^{-11}$ & 9.61$\times\,10^{-11}$ & 17\%\\
\hline
\end{tabular}
\textsc{Notes.}-- $\Phi$ is the space density for objects with $L_ {\nu_{r}=5}>10^{27}$WHz$^{-1}$ in our sample, in sources$/Mpc^3$, $\Phi_{O'Dea}$ is the space density for the O'Dea sample, in sources$/Mpc^3$, $\Phi_{combined}$ is the space density for our and the O'Dea sample combined, in sources$/Mpc^3$, and the error given is a percentage error for the space density for both samples combined, given by $1/\sqrt{N}$.
\end{table}

\begin{figure*}
\begin{center} 
\includegraphics[scale=0.4]{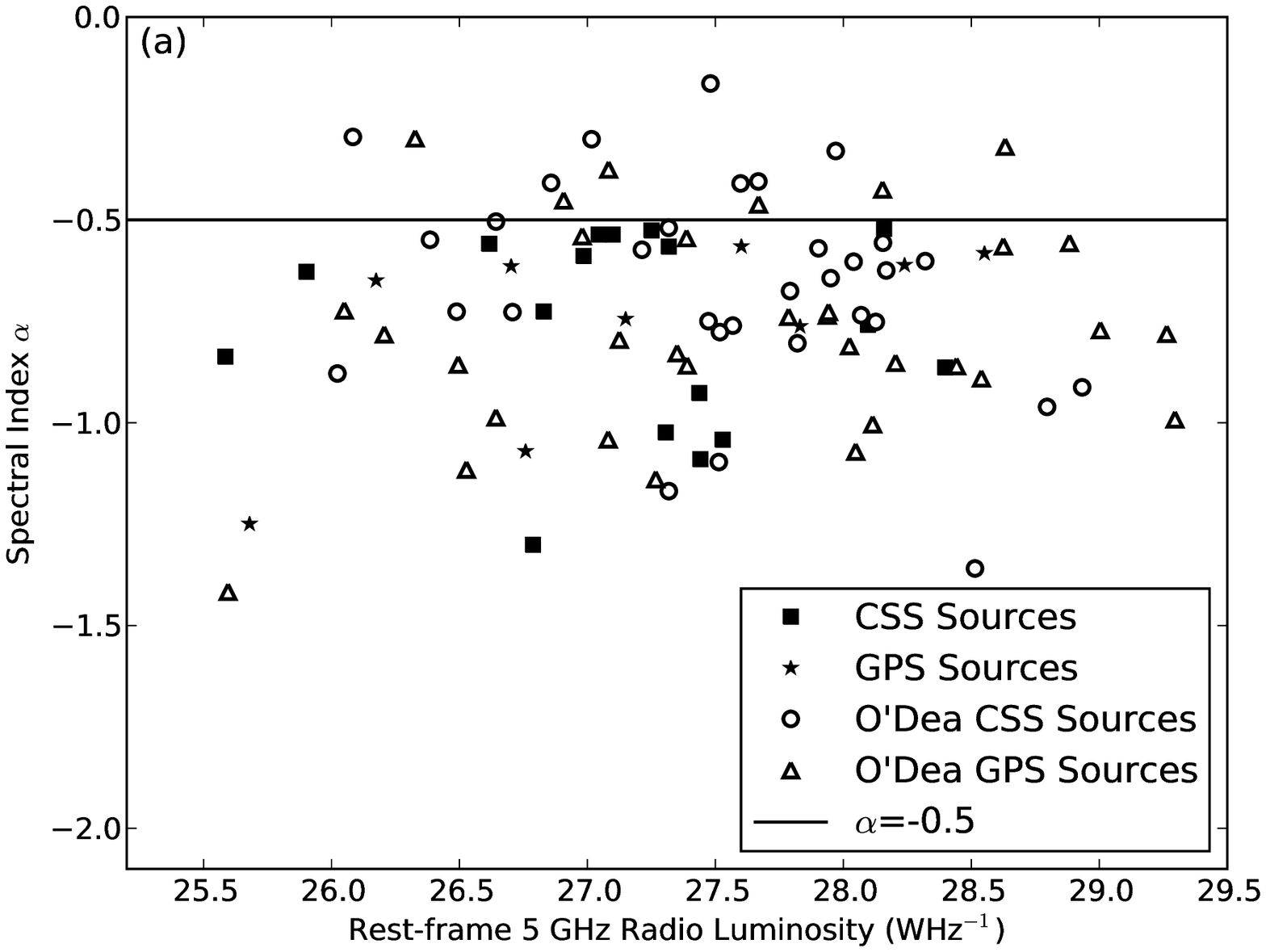}\includegraphics[scale=0.4]{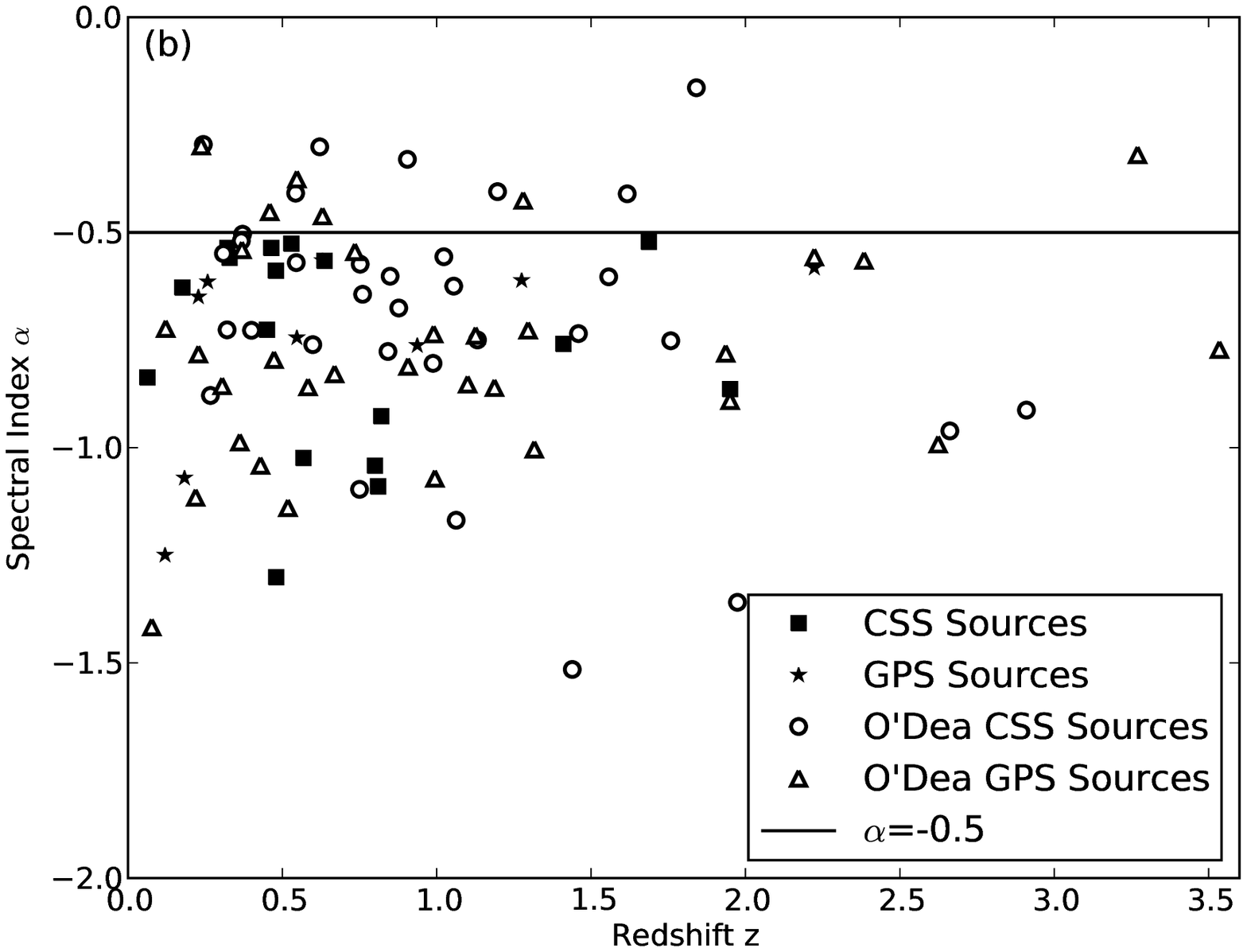}
\caption{(a) Radio rest frame  $\nu_{r}=$5\,GHz luminosity versus spectral index $\alpha$ for our sample and the O'Dea sample. We see several sources from the O'Dea sample have spectral indices $\alpha>-0.5$. (b) Redshift z versus spectral index $\alpha$ for our sample and the O'Dea sample.}
\label{fig:speclum}
\end{center}
\end{figure*}

Figure~\ref{fig:speclum} shows spectral index against both the radio $\nu_{r}=$5\,GHz luminosity and redshift. We can see that several of the O'Dea sources have a spectral index $\alpha>-0.5$, and those that have the highest luminosity are dominated by spectral indices close to $-0.5$. For GPS sources, however, the spectral index may be calculated closer to the peak for some sources than others, skewing the value towards a less steep spectrum. 

In summary, the main differences between our sample and O'Dea's sample are the following:
\begin{itemize}
\item O'Dea's sample selects higher redshift (and hence higher luminosity) objects that we do not see in our sample.
\item Our optical hosts are mostly (61\%) unresolved, yet $>73$\% of the O'Dea sample show either a second nucleus or indications of interactions, distortions and or mergers, although both samples have comparable resolution.
\item Four of the eight sources in the overlap region between our sample and the O'Dea sample do not satisfy our selection criteria. Similarly, other known CSS and GPS sources failed one or more of the selection criteria used in this investigation.
\end{itemize}
Clearly, an algorithm to select a homogeneous flux limited sample is necessary to confirm how these powerful sources fit into the overall evolution of radio galaxies. 

\section{Conclusions}
\label{sec:concl}
We present an unbiased flux limited sample of 26 CSS and GPS objects, consisting of 2 new candidate and 15 known CSS sources and 9 known GPS sources. This represents the first step toward a thorough and complete catalogue of bright CSS and GPS sources in the southern hemisphere. With this sample we can now begin to explore in detail the properties of young AGN, by comparison of this sample to our faint sample, which will be presented in later papers. 
It is clear, however, that there are many differences between the various existing samples of CSS and GPS sources, particularly depending on the selection criteria. Given these discrepancies, the overall properties of this population remain poorly constrained and require more detailed investigation. The selection of these sources is challenging, and we propose here an algorithm for selecting these populations in large numbers. 

Further investigation could include deeper optical imaging for those sources without optical counterparts in SuperCOSMOS, spectroscopy to confirm photometric redshifts and obtain redshifts for those sources without, as well as studying the 50 poor candidates cut from our final catalogue. Fractional polarization measurements would enable comparison with previous CSS and GPS samples. Redshift information allows the linear size of the resolved objects to be measured, which should be very small for CSS or GPS sources, as this is a key feature of these young objects. Redshifts are critical for investigating evolution to provide a robust picture of how this population fits within the broader context of galaxy and AGN evolution. With a complete picture of the bright CSS and GPS population in hand, a thorough comparison to fainter such populations can then be robustly pursued.
	
\section*{Acknowledgements}
The Australia Telescope Compact Array is part of the Australia Telescope which is funded by the Commonwealth of Australia for operation as a National Facility managed by CSIRO.\\
This research has made use of the NASA/IPAC Extragalactic 
Database
(NED), which is operated by the Jet Propulsion Labratory, Caltech, under contract with NASA.\\
We also wish to thank Elaine Sadler for all her time and effort with this paper. Her efforts are greatly appreciated.\\

\section*{Supporting Information}
\noindent
An additional table of Supporting Information is available in the online version of this article. 

\noindent
\textbf{Table~\ref{table:sample}.} The full version of this table, containing all flux density information, optical magnitudes, redshift, physical and angular sizes, and other relevant information.

\noindent
Please note: Wiley-Blackwell are not responsible for the content or functionality of any supporting material supplied by the authors. Any queries (other than missing material) should be directed to the corresponding author for the article.

\bsp

\label{lastpage}

\end{document}